\documentclass[3p,authoryear]{elsarticle}
\usepackage{deluxetable}
\usepackage{subfig}
\usepackage{natbib}	
\usepackage{hyperref} 
	
\title{PRIMASS visits Hilda and Cybele groups}

\author[on]{M. N. De Pr\'a}
\ead{mariodepra@on.br}
\author[noe]{N. Pinilla-Alonso}
\ead{kaveh@river-valley.com}
\author[on]{J. M. Carvano}
\author[iac,ull]{J. Licandro}
\author[camp]{H. Campins}
\author[thais]{T. Moth\'e-Diniz}
\author[iac,ull]{J. De Le\'on}
\author[victor]{V. Al\'i-Lagoa}

\address[on]{Departamento de Astrof\'isica, Observat\'orio Nacional, Rio de Janeiro, 20921-400, Brazil}
\address[noe]{Florida Space Institute, University of Central Florida, Florida, USA}
\address[iac]{Instituto de Astrof\'isica de Canarias, C/V\'ia Láctea s/n, 38205 La Laguna, Spain}
\address[ull]{Departamento de Astrof\'{\i}sica, Universidad de La Laguna, 38206 La Laguna, Tenerife, Spain}
\address[camp]{Department of Physics, University of Central Florida, 4111 Libra Drive, PS 430, Orlando, FL 32826, USA}
\address[thais]{Department of ICT and Natural Sciences, Norwegian University of Science and Technology,  \AA lesund, Norway}
\address[victor]{Max-Planck-Institut f{\"u}r extraterrestrische Physik, Giessenbachstrasse 1, 85748 Garching, Germany}

\journal{Icarus}

\begin{document}
	\begin{abstract}
        The Cybele and Hilda dynamical groups delimit the outer edge of the asteroid belt. Their compositional distribution is a key element to constrain evolutionary models of the Solar System. In this paper, we present a compositional analysis of these populations using spectroscopic observations, SDSS and NEOWISE data. As part of the PRIMASS (Primitive Asteroids Spectroscopic Survey), we acquired visible spectra of 18 objects in Hilda or Cybele groups with the Goodman High Throughput Spectrometer at the 4.1m SOAR telescope and 20 near-IR spectra of Hilda objects with Near Infrared Camera Spectrograph at the 3.56m TNG. The sample is enlarged with spectra taken from the literature in order to increase our statistical analysis. The spectra were inspected for aqueous alteration bands and other spectral features that can be linked to compositional constraints. The analysis shows a continuous distribution of compositions from the main-belt to the Cybele, Hilda and Trojan regions. We also identify a population in the Trojans group not present in Hilda or Cybele objects.  
		
		
	\end{abstract}
	\maketitle
	
	\section{Introduction}
	The outskirts of the asteroid belt can be divided into three main groups: the Cybele, between 3.3 and 3.7 au, in the external region of the Hecuba gap (i.e., the 2:1 mean motion resonance with Jupiter), the Hilda  at $\sim$ 4.0 AU in the 3:2 mean motion resonance with Jupiter, and the Trojan population around the L4 and L5 equilibrium points of Jupiter. 
	
	Due to their heliocentric distances, asteroids belonging to these groups would have experienced less heating and should be of more pristine composition than objects in inner regions of the main belt \citep{rivkin2015,krot2015}. Therefore they are considered to be transitional populations between icy and rocky objects. 
	
	Early investigations of the composition of members of the outer belt populations (Tedesco \& Gradie, Gradie 1989, Gradie 1979) showed a predominance of asteroids with low albedo and featureless spectra, whose colors vary from gray to red. In the Tholen's taxonomic classification \citep{tholen1984asteroid} this corresponds to the C-, P-, or D-type.	The red color in primitive class asteroids is often associated with the presence of complex organics on their surfaces \citep{gaffey1989,vilas1994iron}. Moreover, \cite{emery2006thermal} detected fine grained anhydrous silicates on the surface of D-/P-type Trojan asteroids by their thermal emission. Although these taxonomic classes are observed across the whole asteroid belt, \citet{carvano2003s} pointed out, based on visible spectroscopy, that inner belt D-type objects often have concave spectral shapes and higher albedo compared to the outer belt D-types, suggesting that they may be compositionally different. 

		\begin{table*}[ht]
			\centering
			\caption{Asteroids observational conditions - Visible}
						\small
			\begin{tabular}{llcccccccc}	\hline	\hline
				
				Number & Name & Date & UT  &  Airmass &  $m_v$ & $\alpha$ & T$_{EXP}$ &Slit & SA*\\
				&  & start &   &  &   &(o) & sec & ('') &\\
				
				\hline
				225 & Henrietta & 2011-01-31 & 05:32 & 1.11 & 14.93 & 5.60 & 270.0 & 1.03 & 1 \\
				229 & Adelinda & 2012-03-28 & 00:43 & 1.623 & 14.94 & 6.45 & 330.0 & 1.68 & 1,4,5 \\
				401 & Ottilia & 2011-01-31 & 07:35 & 1.22 & 15.02 & 14.60 & 240.0 & 1.03 & 1 \\
				528 & Rezia & 2012-03-28 & 09:10 & 1.163 & 15.40 & 16.63 & 360.0 & 1.68 & 1,4,5 \\
				790 & Pretoria & 2012-03-29 & 00:05 & 1.257 & 14.57 & 14.63 & 240.0 & 1.68 & 1,3,5 \\
				909 & Ulla & 2011-01-31 & 01:28 & 1.353 & 14.89 & 17.38 & 180.0 & 1.03 & 1 \\
				940 & Kordula & 2011-02-01 & 07:26 & 1.363 & 16.04 & 14.59 & 540.0 & 1.03 & 1 \\
				1177 & Gonnessia & 2012-03-28 & 00:01 & 1.213 & 14.53 & 11.86 & 240.0 & 1.68 & 1,4,5 \\
				1280 & Baillauda & 2012-03-29 & 01:45 & 1.363 & 16.04 & 13.21 & 360.0 & 1.68 & 1,3,5 \\
				6039 & Parmenides & 2011-02-01 & 05:22 & 1.187 & 17.13 & 4.66 & 600.0 & 1.03 & 1 \\
				\hline
				334 & Chicago & 2011-01-31 & 05:13 & 1.51 & 13.11 & 0.12 & 73.333 & 1.03 & 1 \\
				1144 & Oda & 2011-02-08 & 06:11 & 1.387 & 16.41 & 11.84 & 540.0 & 1.03 & 1 \\
				1269 & Rollandia & 2011-02-01 & 03:48 & 1.603 & 14.14 & 5.45 & 180.0 & 1.03 & 1 \\
				1439 & Vogtia & 2012-03-28 & 03:26 & 1.253 & 15.73 & 4.21 & 360.0 & 1.68 & 1,4,5 \\
				1902 & Shaposhnikov & 2012-03-27 & 06:46 & 1.193 & 15.72 & 14.30 & 240.0 & 1.68 & 1,2,3,4 \\
				3202 & Graff & 2011-02-01 & 01:02 & 1.338 & 16.43 & 10.79 & 510.0 & 1.03 & 1 \\
				3577 & Putilin & 2011-02-01 & 04:10 & 1.49 & 14.98 & 2.81 & 180.0 & 1.03 & 1 \\
				3843 & Oisca & 2011-01-31 & 04:39 & 1.75 & 16.48 & 5.91 & 480.0 & 1.03 & 1 \\
				7394 & Xanthomalitia & 2012-03-29 & 00:33 & 1.373 & 17.52 & 12.35 & 600.0 & 1.68 & 1,3,5 \\
				\hline
				\multicolumn{9}{l}{\footnotesize *Solar Analogs: (1) L102-1081, (2) L107-684, (3) L107-998, (4) HD44594, (5) HD144584}
				
			\end{tabular}
			\label{table:visobs}
		\end{table*}

	Furthermore, \citet{dahlgren1995study} $\hspace{0.026cm}$ and  \citet{dahlgren1997s} investigated visible spectra of 43 objects in the Hilda group.  They reported 64\% of the Hilda asteroids belonged to the D-class, while 28\% and 2\% were P- and C-types, respectively (the remaining percentage belonged to ambiguous classes). In addition, a relation between spectral slope and asteroid size was found. The authors argued that this could be the result of a size dependent surface composition where the P-types dominate at larger sizes. A possible explanation is given by their mutual collisions, if D-types are more fragile than P-types, this will favor disruptive collisions among D-type precursors. In this case, a larger fraction of the smaller body population can be collisional fragments from a few shattered large D-type precursors resulting in a large fraction of small D-type asteroids, as observed.

	Investigations on Cybele asteroids composition have been carried out by \citet{lagerkvist2005s}. They obtained visible spectra of 20 Cybele asteroids and found that the D-type Cybele objects tend to be smaller than P- and C-type objects, which is similar to the aforementioned behavior for the Hilda group. Additionally, they note the presence of one large S-type among the Cybele objects and a larger fraction of C-types than in Hilda population.

	The results of \citet{dahlgren1997s} were obtained using reflectance spectra of asteroids with absolute magnitude $H_V < 11.3$, which means diameters $D > 35$ km assuming an albedo of $p_V = 0.05$, and the results of \citet{lagerkvist2005s} were obtained using reflectance spectra of Cybele asteroids with absolute magnitude $H_V < 11.9$, which means  $D > 20$ km assuming the same $p_V$. Both samples correspond only to the large end of the size distribution of the Cybele and Hilda asteroids.
	
    \citet{gil2008surface} and \citet{gil2010tax} searched for photometric data of Hilda and Cybele asteroids, respectively, in the Moving Object Catalogue of the Sloan Digital Sky Survey to find the spectrophotometric characteristics of small members of both groups. They found that the correlation between size and spectral slope previously suggested for Hilda and Cybele asteroids was correct only for large objects ($H < 12$) but it was not supported by data obtained from the small ones. The authors propose that the observed trend could be the result of a combination of the space weathering and resurfacing due to a collisional process modified by a truncation of the population size distribution.
    
    While several tens of visible spectra of Cybele and Hilda asteroids have been published, there are only a few of them in the near-infrared region. \citet{dumas1998near} reported spectra of 1 Cybele and 8 Hilda asteroids in the 0.8-2.5 $\mu$m spectral region together with the spectra of another 9 low albedo asteroids. The selected targets belonged to the P- or D-types in the taxonomy classification of \cite{tholen1984asteroid}, all objects presented slightly red and featureless spectra. 

    Recently, \citet{takir2012outer} published spectra of 6 Cybele and 3 Hilda asteroids among 28 primitive asteroids with a semi-major axis of $2.5 - 4.0 au$ covering the $0.5-4.0 \mu$m region, aiming to examine the distribution and abundance of hydrated minerals (any mineral that contains $H_2O$ or $OH$ associated). They identified four groups on the basis of the shape and band center of the 3 $\mu$m feature: (1)  the "sharp"  group, that exhibits a sharp 3 $\mu$m feature, attributed to hydrated minerals (phyllosilicates); (2) the "Ceres-like" group, that like asteroid Ceres, exhibits a 3 $\mu$m feature with a band center of $\sim3.05$ which is superimposed on a broader absorption feature from 2.8 to 3.7 $\mu$m; (3) the "Europa-like" group, that exhibits a 3 $\mu$m feature with a band center of 3.15 $\pm$ 0.01 $\mu$m; (4) the "rounded" group, that are characterized by a rounded shape feature, attributed to $H_2O$ ice already identified in the infrared spectra of (24) Themis \citep{campins2010themis,rivkin2010}, (65) Cybele \citep{licandro2011cybeleice} and (107) Camilla \citep{hargrove2012asteroids}. Unlike the sharp group, the rounded group did not experience aqueous alteration. In the Cybele group, five out of six objects presented a 3.0 $\mu$m band that were classified in the "rounded" group, only one belonged to the "sharp" group. While in the Hildas there were three in the "rounded" group and one in the "sharp" group.

		\begin{table*}[!ht]
			\centering
			\caption{Asteroids observational conditions - IR}
			\small
			\begin{tabular}{cccccccccc} \hline \hline
				Number & Name & Date & UT  &  Airmass &  $m_v$ & $\alpha$& T$_{EXP}$ & Slit &SA*\\
				&  & start &  &   &   &(o) & sec & ('') & \\
				\hline
				190	  & Ismene     & 2001-08-04 & 01:00:32 & 2.0 & 14.6 & 10.5  & 30 x 4 & 1.5  & 5    \\
				&            & 2001-09-01 & 22:21:33 &  1.2 & 15.2 & 11.1  & 60 x 8 &1.5  & 2   \\	  
				334	  & Chicago    & 2001-08-05 & 06:45:30 &  1.3 & 14.6 & 13.7  & 50 x 4 &1.5  & 2,5,4   \\
				1202 & Marina     & 2001-08-05 & 04:46:41 &  1.5 & 14.9 &  5.7  & 30 x 8 &1.5  & 2,5,4    \\
				1269  & Rollandia  & 2001-08-05 & 06:30:06 &  1.1 & 15.7 & 14.2  & 50 x 4 &1.5  & 2,5,4    \\
				1754  & Cunningham & 2001-08-05 & 05:29:15 &  1.1 & 15.8 & 12.6  & 50 x 4 &1.5  & 2,5,4     \\
				2067  & Aksnes     & 2001-08-31 & 21:51:18 &  1.6 & 17.8 & 12.2  & 60 x 8 &1.5  & 2    \\
				&            & 2001-09-01 195& 21:57:56 &  1.6 & 17.8 & 12.3  & 60 x 4 &1.5  & 2    \\
				2624  & Samitchell & 2001-09-29 & 01:06:48 &  1.3 & 16.2 &  7.8  & 60 x 4 &1.5  & 2     \\
				3557  & Sokolsky & 2001-08-05 & 05:07:45 &  1.1 & 16.5 & 18.9  & 50 x 4 &1.5  & 2,5,4  \\
				3561  & Devine     & 2001-08-05 & 05:50:33 &  1.2 & 16.9 & 12.8  & 50 x 4 &1.5  & 2,5,4  \\
				4317  & Garibaldi  & 2001-09-30 & 03:15:38 &  1.1 & 17.0 &  9.5  & 60 x 8 &1.5  & 2,5,4   \\
				5368  & Vitagliano & 2001-09-30 & 03:43:34 &  1.2 & 17.3 & 12.8  & 60 x 11&1.5  & 2,5,4   \\
				5661  & Hildebrand & 2001-09-29 & 23:43:08 &  1.2 & 15.9 &  6.4  & 60 x 41&1.5  & 2,5,4    \\
				5711  & Eneev      & 2001-09-29 & 01:26:39 &  1.3 & 16.1 &  7.0  & 60 x 10&1.5  & 2    \\
				6237  & Chikushi   & 2001-10-05 & 23:12:22 &  1.2 & 17.3  & 3.6  & 60 x 8 &1.5  & 2,4,3   \\
				9121  & Stefanovalentini   & 2001-10-05 & 20:32:34 & 1.4 & 17.1 & 10.8 & 60 x 8  &1.5  &2,4,3    \\
				11750 &            & 2001-10-06 & 05:41:49 & 1.2 & 18.5  &  8.8 & 60 x 32 &1.5  & 2,4,3    \\
				15417 & Babylon    & 2001-10-05 & 21:37:36 & 1.4 & 17.7  &  9.3  & 60 x 20 &1.5  & 2,4,3    \\
				15505&        & 2001-10-06 & 04:55:42  & 1.0 & 18.1  & 9.5   & 60 x 24 &1.5  &2,4,3     \\
				15540 &    & 2002-04-26 & 05:36:33 &  1.2 & 17.9  & 12.6   & 60 x 8  &1.5  & 1,2,5   \\
				\hline
				\multicolumn{9}{l}{\footnotesize *Solar Analogs: (1) L102-1081, (2) L110-361 (3) L98-978, (4)L93-101, (5) 112-1333}
			\end{tabular}
			\label{table:irobs}
		\end{table*}
	
 	Even though the spectral analysis of the 3.0 $\mu$m region can provide a clue to the presence of water ice or hydrated minerals on a primitive asteroid surface, in the visible and near-infrared regions (up to 2.4 $\mu$m)  the lack of specific spectral features prevents an unique compositional interpretation. Even if weak minor absorption bands has been reported in these regions, their interpretation is not clear \citep{thais}. At present, the general outline for the composition of these asteroids is a mixture of organics, anhydrous silicates, opaque materials and ice \citep{bell1989m,gaffey1989,vilas1994iron}. It is very difficult to define the composition of these objects since no analogous meteorites for P-type asteroids has been found and there is only one analogous meteorite for D-types: the Tagish Lake, a very red and opaque meteorite \citep{hiroi2001tagish}.	
	
    Considerable interest in studying the Hilda and Cybele populations is also due to their possible relation with dormant comets \citep{licandro2008spectral}. \cite{disisto2005hildacomet} shows that a considerable amount of the Jupiter family comets could have actually been originated from the Hilda population.
    
    Planetary migration models, such as the Nice model \citep{gomes2005origin,morbidelli2005,tsiganis2005origin}, posit that a strong dynamical evolution would have occurred in the early Solar System, mainly due to interactions between Jupiter and Saturn. In particular, the Hilda and Cybele populations would be directly affected by the orbital configuration and evolution of the giant planets.  Such a scenario would destabilize the Jovian Trojan and Hilda populations, repopulating them later during the same phase of the dynamical evolution with planetesimals scattered inward from the region beyond the ice giants \citep{gomes2005origin,roig2015evolution,morbidelli2005,brovz2008asteroid}. The Cybele asteroids are the last stable region of the main belt, before the resonant populations. \citet{levison1993gravitational} showed that some objects originating in the primitive trans-Neptunian belt may have also been inserted in the outer regions of the main belt, during the same period. From these models, therefore it is expected that Hilda and Trojan's populations show a similar compositional distribution, while the Cybele group should present a broader distribution of surfaces since it would have objects with origin both in the main belt and in the trans-Neptunian belt.

	In this paper we present new spectroscopic data of Cybele and Hilda asteroids in the visible and near-IR. We also analyze the visible and near-infrared (near-IR) spectra in the literature. In section~\ref{sec:obs} we describe the observation and reduction processes, and in section \ref{sec:analysis} the parametrization and analysis. The results for the spectroscopic analysis are presented in section \ref{sec:results}. In section \ref{sec:sdss} we perform an extended analysis using data of Hilda and Cybele groups available in large public databases, such as SDSS and NEOWISE. The discussion of our results is presented in section \ref{sec:discussion}, and finally the conclusions on section \ref{sec:conclusions}.
	

	\begin{figure}[!ht]
		\begin{center}
			\includegraphics[angle=0,scale=.72]{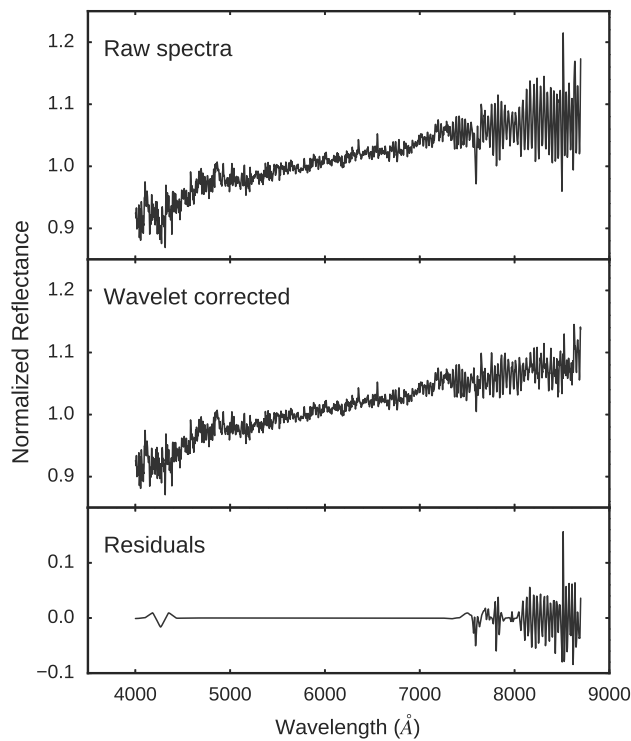} 
		\end{center}
		\vspace{-0.5cm}
		\caption{Result from wavelet technique to filter out fringing effects from the spectra using grating 300 l/mm. Top image shows the spectrum of (790) Pretoria; Middle image shows the spectrum after wavelet filtering and bottom image the residuals.} 
		\label{fig:wav} 
	\end{figure}

	\section{Observation and data reduction}
	\label{sec:obs}
	\subsection{Visible}
		\label{subsec:obs}
	We collected low-resolution spectra of 18 asteroids in the Cybele and Hilda populations (Table~\ref{table:visobs}). The data were obtained through the Goodman High Throughput Spectrograph (GTHS) at the 4.1m SOAR telescope on Cerro-Pach\'on, Chile. We used a setup with the grating of 300 $lines/mm$ and the slits of 1.03'' in 2011 and 1.68'' in 2012 with no second order blocking filter, which provides an effective spectral interval of 0.4-0.87 $\mu$m. Observations were made in a total of 6 nights, split in semesters 2011A and 2012B. We also obtained two sequences of calibration quartz lamps, just before and after the acquisition of the target. Acquiring them with the same configuration as the target enabled us to account for flexures of the instrument. At least one solar analog was observed during each night, at different airmasses. 
 
	The quartz lamps were used to do the flat-field correction of the images, while the HgAr lamps were used for the wavelength calibration. We applied standard reduction techniques: images were bias and flat-field corrected using quartz lamp flat. In sequence, the sky background was subtracted and each one-dimensional spectrum was extracted with variable aperture, depending on the conditions of the night. The spectra were wavelength calibrated with HgAr lamps. This procedure was repeated for the three sub-exposures of each target. The spectra were then averaged to produce a final object spectrum.

	To obtain asteroid reflectance spectra, we divided the object spectrum by the spectra of Solar Analogs. Before comparing the spectra of the target with the spectra of the solar analogs to remove the signature of the Sun, we analyzed the spectra of the standard stars to detect small differences in color introduced during the observations, e.g. by inconsistent centering of the star in the slit. These differences could propagate into the spectrum of the target through the reduction process. To quantify these errors,  we divided, for each night, all of the spectra of the solar analogs by one that we take as reference (the one at lower airmass), after applying an atmospheric extinction correction.	 The extinction of the sky is dependent on the wavelength, with shorter wavelengths experiencing greater extinction. To minimize the spectral extinction effect from the difference in airmass between the stars and the target, we applied color correction to the spectra of the object and the stars. In the absence of extinction coefficients for Cerro Pach\'on, we used the mean extinction coefficients for La Silla, since this observatory is located relatively close and at similar altitude from Cerro Pach\'on. A study of the variation of extinction coefficients from different sites suggested the extinction is mostly influenced by the altitude of the site. The result of dividing the spectrum of a solar analogs by another should be a straight line with spectral slope $S' $ = 0. This procedure enable us to discard observations of stars with a bad behavior induced by systematic errors and estimate the error in the slope. Finally, all reflectance spectra were normalized to 1 at 0.55$\mu$m.
		
	The data reduction was made by combining scientific Python with IRAF\footnote{IRAF is distributed by the National Optical Astronomy Observatories, which are operated by the Association of Universities for Research in Astronomy, Inc., under cooperative agreement with the National Science Foundation.} tasks, called through the PyRAF\footnote{PyRAF is a product of the Space Telescope Science Institute, which is operated by AURA for NASA.} library.

	\subsubsection{Fringing correction}
	
	The final spectra presented a strong fringing pattern towards the red part of the spectrum. The pattern is still noticeable even after the flat field correction. In order to attenuate this fringing contribution, we applied a wavelet technique based on \cite{mallat}. This type of algorithm is typically used for signal denoising, i.e. decreasing the intensity of high frequencies in the wavelet decomposition. Our approach was to establish a bandpass algorithm that decrease intensity of medium-high frequencies, without removing the high frequencies (noise). We applied a coiflet wavelet with hard thresholding with an up and low threshold. A typical result is shown on Figure \ref{fig:wav}.
	
	\subsection{Near-infrared}
	\label{sec:nearirobs}

	Low  resolution  near-infrared  spectra  were  taken  with  the 3.56 m Telescopio Nazionale Galileo (TNG) using the low resolution  mode  of  NICS  (Near  Infrared  Camera  Spectrograph), based on an Amici prism disperser that covers the 0.8-2.4 $\mu$m region (Oliva 2000). The slit was oriented in the parallactic angle, and we used differential tracking that follows the asteroid motion. The width of the slit used was 1.5'' and corresponds to a spectral resolving power R $\sim$ 34 quasi-constant along the spectra. The observational method and reduction procedure followed that described in Licandro et al. (2002a).
	
	\begin{figure*}[!ht]
		\centering
		\begin{tabular}{cc}
			{\includegraphics[angle=0,scale=.5]{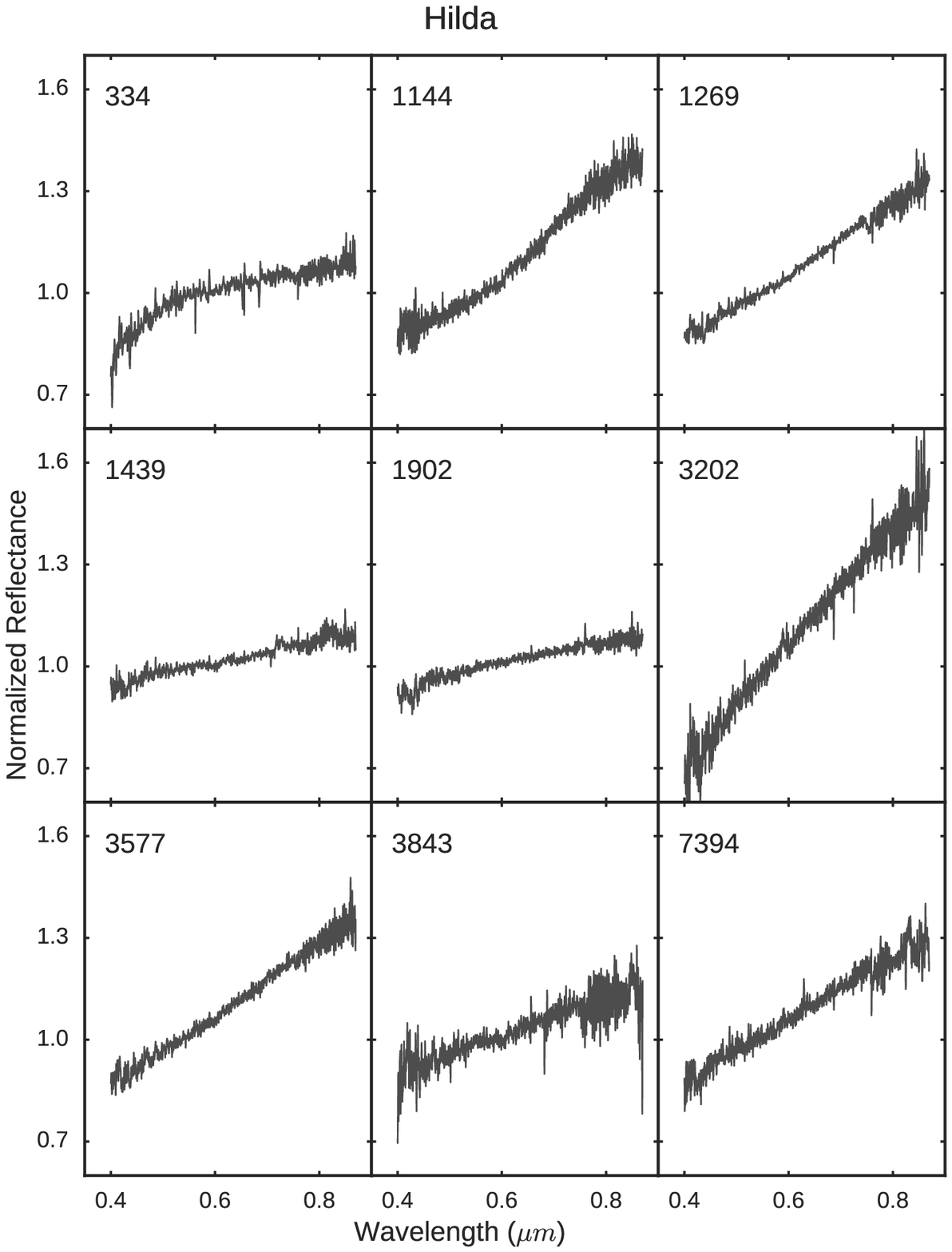}
			}
			{\includegraphics[angle=0,scale=.5]{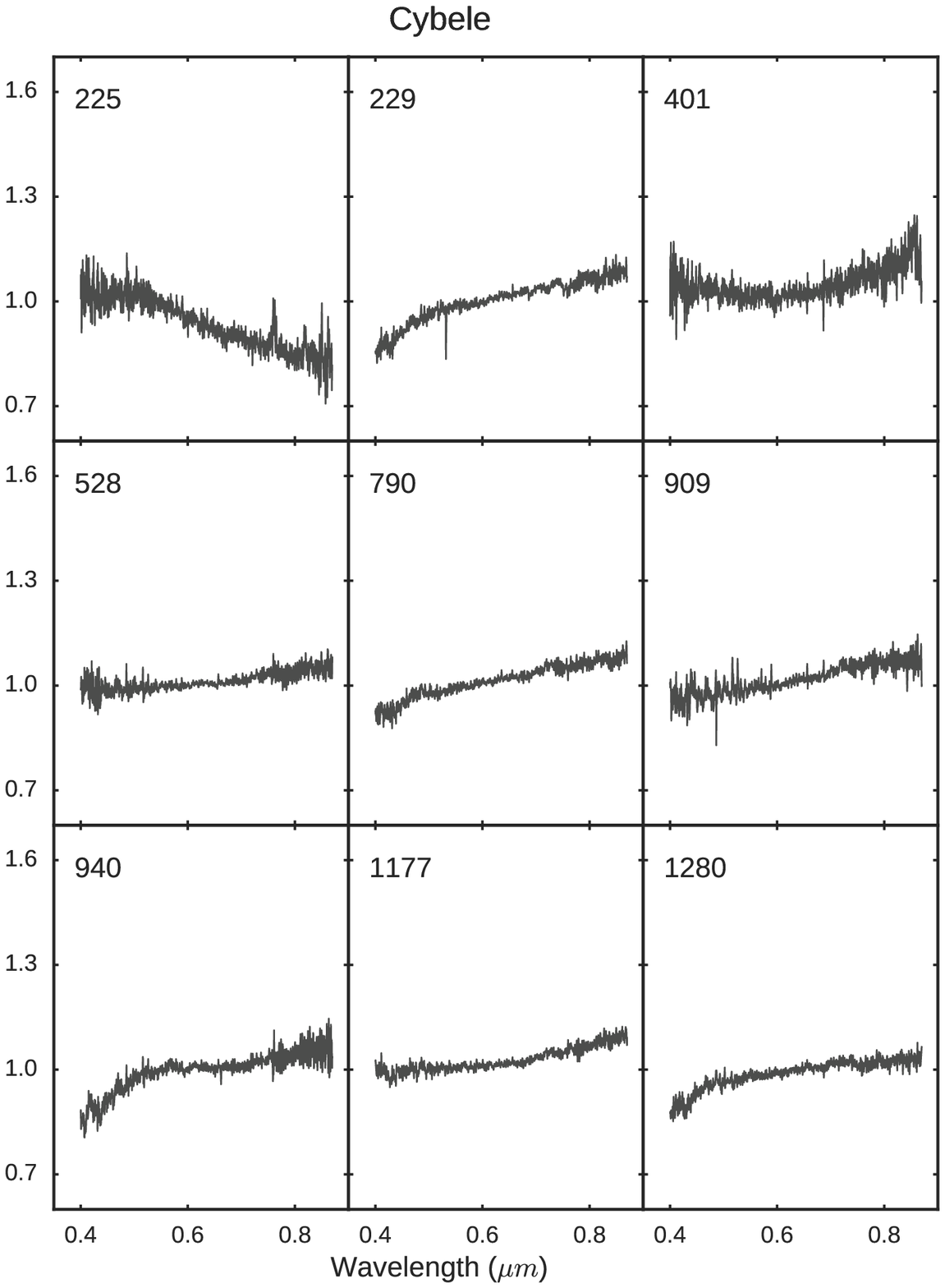} }
		\end{tabular}
		
		\caption{Visible spectra of Hilda and Cybele asteroids, acquired with GHTS-SOAR during the campaigns presented in this work. All spectra were normalized to unity at 0.55 $\mu$m}	
		\label{fig:specs}
	\end{figure*}

	The acquisition consisted of a series of short exposure images in one position of the slit (position A) and then offsetting the telescope by 10'' in the direction of the slit (position B), and obtaining another series of images. This process was repeated and a number of ABBA cycles were acquired. The total on-object exposure time is listed in Table ~\ref{table:irobs}. The two-dimensional spectra were extracted, and collapsed to one dimension. The wavelength calibration was performed using a look-up table which is based on the theoretical dispersion predicted by ray-tracing and adjusted to best fit the observed spectra of calibration sources and telluric absorptions. To correct for telluric absorption and to obtain the relative reflectance, several G2 stars from the list of Landolt (1992) were observed during the same night at airmass similar to that of the asteroids.  These Landolt stars have been observed on  previous  nights  together with the solar analogue star P330E (Colina \& Bohlin 1997) and they are intensively used as solar analogs.

	Finally, the spectra of the asteroids were divided by the spectra of the solar analogue stars, and the so obtained reflectance spectra averaged, obtaining the final reflectance spectrum of  each object. Sub-pixel offsetting was applied when dividing the two spectra to correct for errors in  the wavelength calibrations due to instrumental flexure. By comparing the reflectance spectra of the same asteroid obtained with different solar analogues we determined that the uncertainty in the slope is smaller than 1\%/0.1 $\mu$m.

	\begin{figure*}[!ht]
		\centering
		\begin{tabular}{cc}
			\subfloat[]
			{\includegraphics[angle=0,scale=.65]{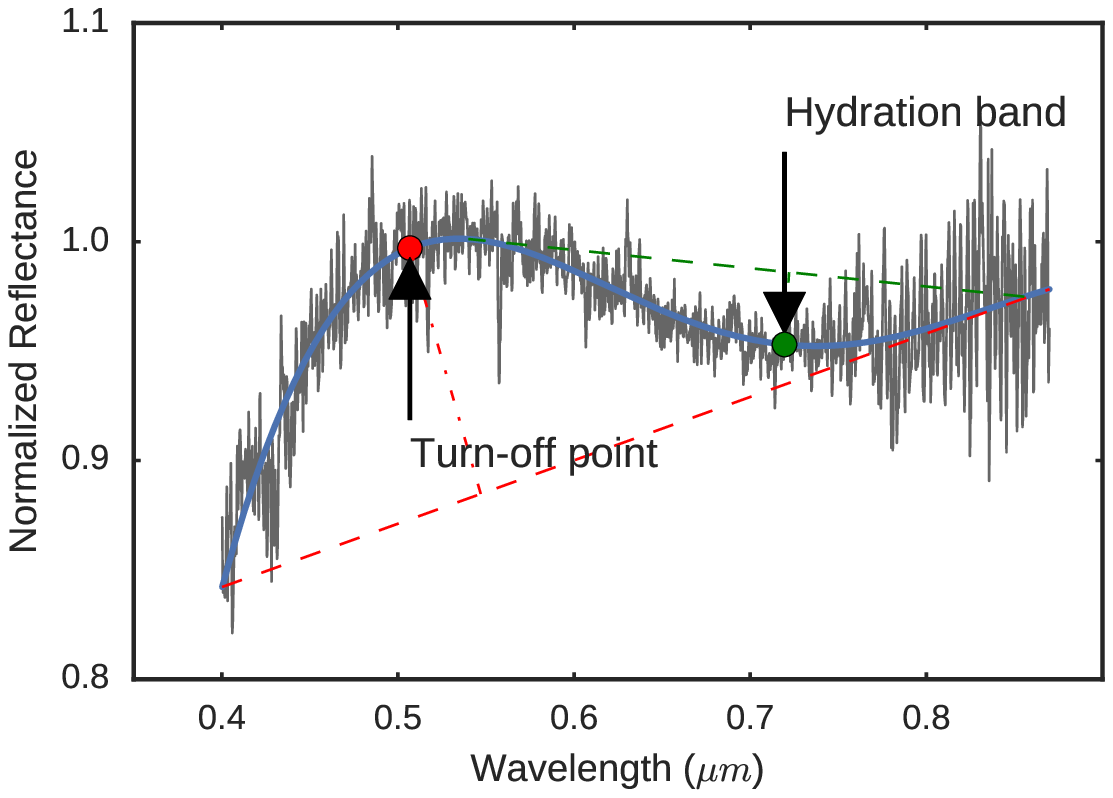}}
			\subfloat[]
			{\includegraphics[angle=0,scale=.65]{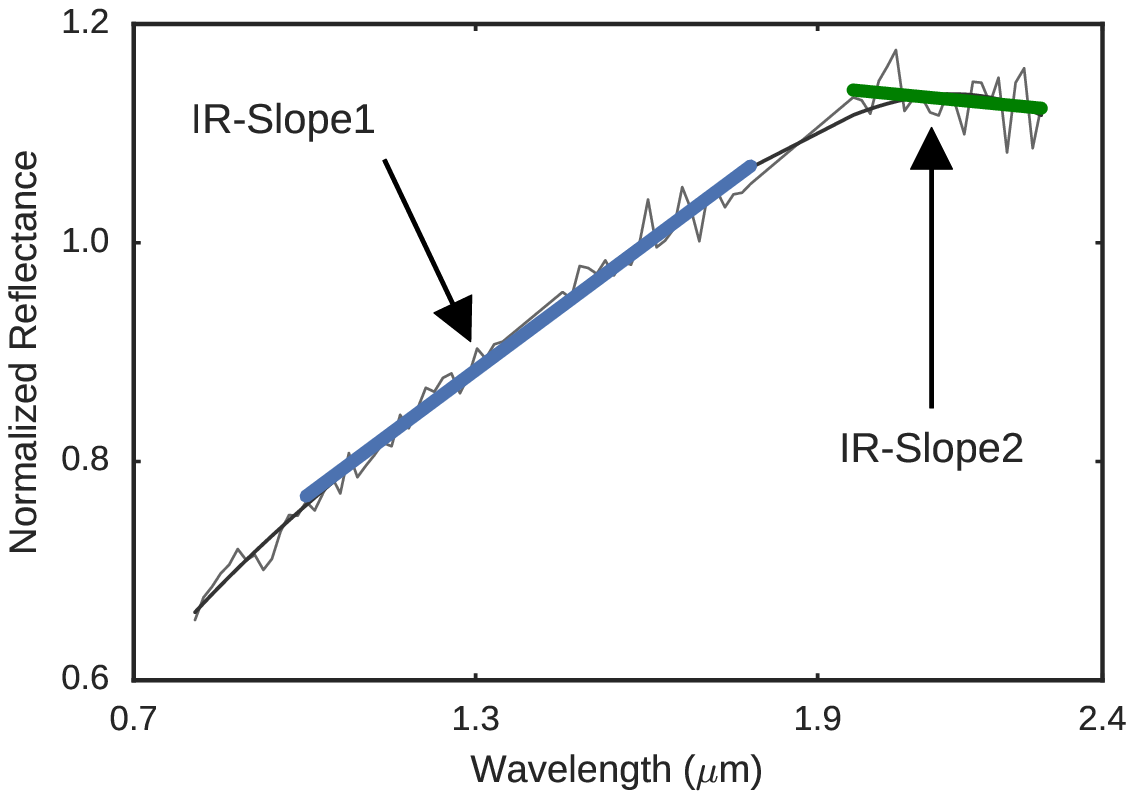} }
			
		\end{tabular}
		
		\caption{(a) - Parametrization for visible spectra in asteroid (940) Kordula. We measure four features: Visible Slope, UV-slope, Turn-off point (red lines) and the Hydration band (green lines)  (b) - Parametrization schema for near-IR spectra in asteroid (1269) Rollandia. Some spectra present a turn in the slope towards the red. We identify this behavior by measuring two slopes: IR-slope1 is measured in the 1.0-1.75 $\mu$m interval, while IR-slope2 in the 1.95-2.3 $\mu$m. }	
		\label{fig:params}
	\end{figure*}
		
		
\section{Analysis}
\label{sec:analysis}

	\subsection{Visible}
	\label{subsec:visible}
	
		We present nine spectra of Hilda asteroids and nine spectra of Cybele asteroids (Figure \ref{fig:specs}). All of them (but one) are red, only (225) Henrietta shows a blue spectral slope. The majority of them are featureless, only (940) Kordula shows a broad absorption band centered at ~0.7 $\mu$m (Figure \ref{fig:params}). While, Some of them also show a clear drop of reflectance bellow 0.5 $\mu$m.	
		
		For the characterization of these spectra we defined 4 parameters(Figure \ref{fig:params}): the presence of a 0.7 $\mu$m absorption band, the visible slope, the presence of a turn-off point around 0.5 $\mu$m and, in case of the existence of the turn-off point, we also measure the near-ultraviolet (near-UV) slope. We also determined the taxonomic classification of these objects in the \cite{bus2002tax} scheme.
		
		In order to increase the sample for the statistical analysis of the populations we collected visible spectra of Cybele and Hilda objects in the literature. From the spectroscopic surveys S3OS2 \citep{lazzaro2004s30s2}, SMASS \citep{bus2002tax} and Vilas et al. (1998), we gathered:  5 spectra from each dataset for the Hilda group, and 35,11,15 for the Cybele group, respectively. We also added 30 spectra from \citet{dahlgren1997s} and \citet{dahlgren1995study} for the Hilda population; and 18 spectra from \citet{lagerkvist2005s} for the Cybele population. Therefore, the total sample of visible data consists in 88 spectra of 55 objects for the Cybele population and 54 spectra of 37 objects for the Hilda population. It is important to note that the spectral coverage of these works are slightly different from the one obtained with GHTS-SOAR; S30S2 has a spectral coverage of 0.5-0.9 $\mu$m; SMASS of 0.4-0.9 $\mu$m; Vilas 0.5-0.9 $\mu$m; \citet{dahlgren1997s} and  \citet{dahlgren1995study}; 0.4-0.9 $\mu$m or 0.4-0.7 $\mu$m, and \citet{lagerkvist2005s} of 0.4-0.9$\mu$m.

		All literature spectra were re-analyzed with the aforementioned parametrization, for the sake of homogeneity, although due to the varying spectral coverage, there are cases where some of the parameters could not be measured.

			\begin{figure*}[!ht]
				\centering
				\includegraphics[angle=0,scale=.5]{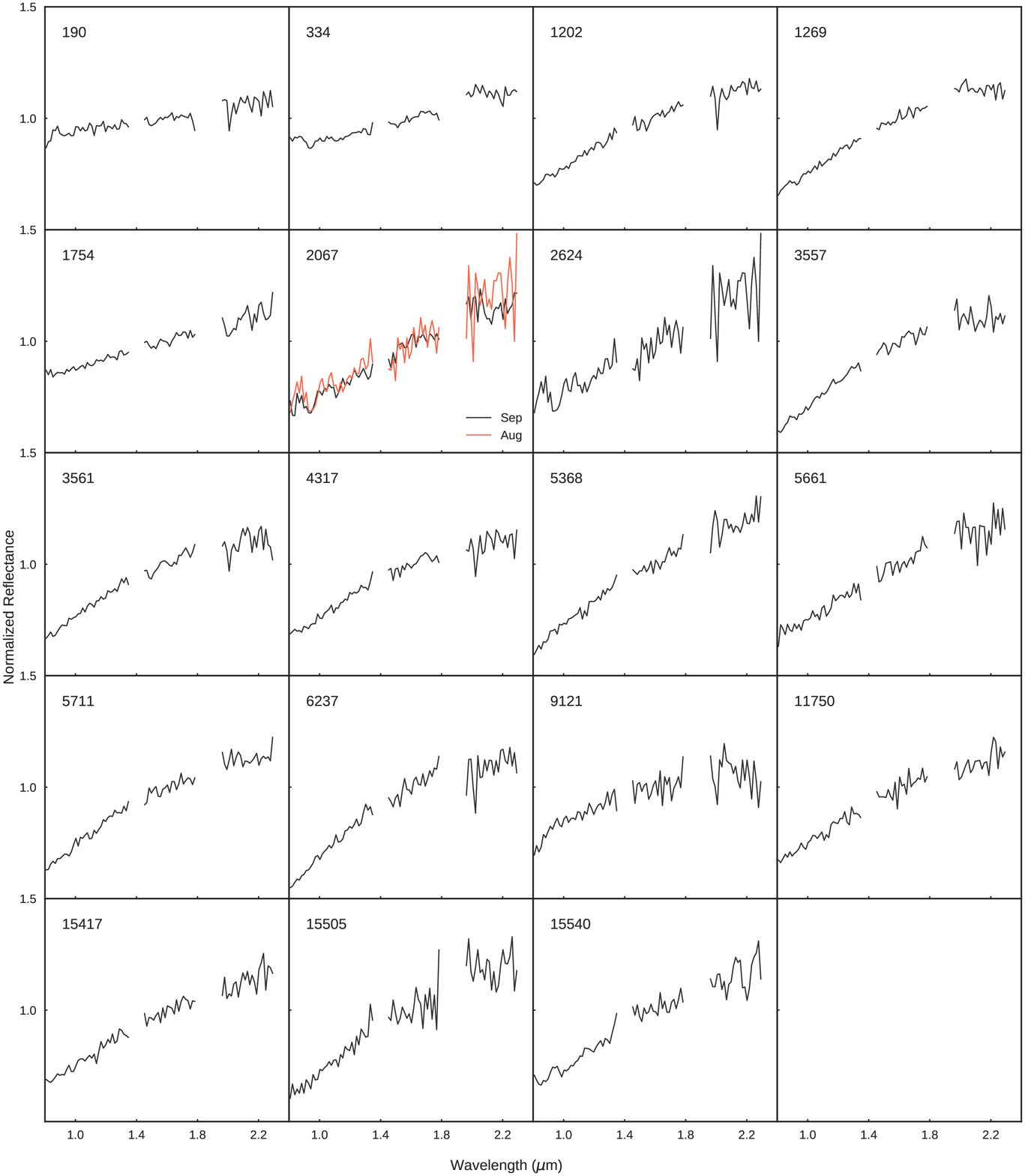} 
				\caption{Near-IR spectra of 20 Hilda objects, observed at 3.6m TNG. Some of the objects were observed more than once.}
				\label{fig:specsir}
			\end{figure*}

	\subsubsection{Taxonomy}
		\label{subsubsec:tax}
		The taxonomic classification was made using the on-line tool for modeling spectra of asteroids, M4AST \citep{popescu2012mast}. We first performed a polynomial adjust, with varying order, that represents the spectrum of the asteroid. Then, the tool compared this fit to templates of each class defined by the \cite{demeo2009tax} taxonomy at the corresponding wavelengths. The adopted taxonomic class is the one with the smallest chi-squared. Then we checked the classification of all objects and for the specific targets where the taxonomic result is related to one feature that does not appear in the wavelength range of study, we reclassified them in the \cite{bus2002tax} taxonomy in cases where the result was an specific class of the \cite{demeo2009tax}.

	\subsubsection{Visible slope}
		\label{subsubsec:slope}
		Given that these are primitive asteroids, the majority of them are featureless, i.e. they show no absorption bands in their visible spectra. The only feature that can be measured in all asteroids of this sample is the visible spectral slope.
		
		To calculate the slope, we follow the definition of the spectral gradient (S') in \cite{jewitt2002slope}. We make a liner fit to the spectrum in the wavelength range 0.55 - 0.86 $\mu$m, where the reflectance is well represented by a linear fit. We normalize the fit by the mean value of the reflectance in the adopted range, in units of  \%/1000\AA. 
		
		The systematic error in the slope is estimated by the standard deviation of the distribution of slopes calculated in the solar analogs analysis, as explained in section \ref{subsec:obs}. Unfortunately, for the majority of the spectra in the literature, we have no information on the systematic error associated with the solar analogs. In this case, we assumed a systematic error of 1$\%$. Another source of error is the computation of the slope. To estimate this, we performed a  Monte-Carlo model to fit the slope of the asteroid, by doing a thousand iterations, and removing randomly 20\% of the points, the error assumed is the one-sigma of the distribution. The final error of the slope, included in table \ref{table:vissoar} is the quadratic sum of the systematic and the Monte-Carlo produced errors, and is strongly dominated by the systematic one, specially with a high signal to noise regime. Whenever an asteroid is observed more than once, we take the mean value for the slope and the error is chosen between the standard deviation of the mean or the error propagation, whichever is higher.

	\subsubsection{Hydration: The 0.7 $\mu$m band and the turn-off point}
		\label{subsubsec:hyd}
		
		The presence of aqueously altered minerals on asteroid surfaces can be inferred by the presence of a shallow absorption band centered at  0.7$\mu$m (See figure 3, panel a). This band is strongly correlated with the unambiguous hydration indicator, the 3.0 $\mu$m absorption band \citep{fornasier2014hydration,vilas1994iron}.

		We search for the presence of this feature in our sample applying a methodology similar to the one in \citet{carvano2003s} and \cite{morate}. First we calculate the continuum with a linear fit within the 0.55-0.58 and 0.83-0.86 $\mu$m intervals. We then divide the spectra by the continuum, and fit a fourth-order spline in the 0.58-0.83 $\mu$m range. For objects that present the feature, we characterize its depth and central wavelength. It is important to note that the aqueous alteration absorption band is centered around 0.7 $\mu$m. Objects with spectra like (401) or (1144) present a concave spectra (Figure \ref{fig:specs}), that could be explained by the presence of an absorption band, but not centered near 0.7 $\mu$m, but, at much lower wavelengths, where no hydration band is expected. We therefore do not include these objects in the list of aqueously altered asteroids. 
		
		For the error estimation we ran a Monte-Carlo model with 1000 iterations, randomly removing 20\% of the points, and measured the band depth at each iteration. The final value for the band depth is the center of the resultant distribution and the error is the variance.  
		
		Another possible indicator of hydration is a decrease in reflectance shortward of 0.5 $\mu$m. \cite{vilas1995turn} states that the reflectance spectra of asteroids believed to contain iron-bearing silicates in their surface materials show a strong UV absorption feature believed to be caused by a a ferric oxide intervalence charge transfer transition (IVCT) centered in the UV. C-, B-, and G- generally exhibit a spectral turnover near 0.5 $\mu$m. It is believed that the presence of opaque materials in the surface of the low-albedo asteroids masks this IVCT in the 0.5-0.75 $\mu$m region and slightly lowers the absorption in the blue/UV spectral region.
				
		We will refer to the presence of this feature by the characterization of the "turn-off point". First we perform a linear fit using just ten points at the beginning and at the end of the spectral coverage. Then we measure the distance between the spectral points and the fit over all the spectral coverage (red line in Figure \ref{fig:params}-a). The distance of the farthest away point should surpass a minimal value of 3.5\% to consider the turn-off. This threshold was defined by trial and error and visual analysis.

	\subsection{Near-Infrared}
		\label{subsec:ir}
		Figure \ref{fig:specsir} shows the 20 near-IR spectra of 19 Hilda asteroids. All objects present a red near-IR spectra with no strong absorption, except for a slope change towards redder wavelengths in some cases, e.g. (1269), (2624) and others. A similar behavior was observed in the near-infrared spectrum of the meteorite Alias \citep{cloutis2011-ci}, a CI-class meteorite, and in a few CM-class meteorites \citep{cloutis2011-cm}. The authors explained the feature with the presence of the mineral Berthierine, a phyllosilicate from the serpentine group. 
			
		The same methodology described in sections~\ref{subsubsec:tax} and \ref{subsubsec:slope} were used for the taxonomic classification and slope calculation. We choose to calculate the near-IR slope in two separate intervals: IR-slope1 in the 1.0-1.75 $\mu$m range; IR-slope2 in the 1.95-2.3 $\mu$m range. Objects with slope variation higher than $1.5$ \%/1000\AA~  are considered to present a turn-off in the near-infrared spectrum, around 1.9 $\mu$m.  We excluded the 1.35-1.45 and 1.75-1.95 $\mu$m regions, due to the strong noise caused by Earth's atmosphere absorption.
	
		We searched the literature in order to increase our sample and extend the near-IR analysis to the Cybele group. We collected 4 spectra from SMASSII, 9 from \cite{takir2012outer} and 2 from Reddy et al (2016). The final sample contains an amount of 40 spectra for 31 objects in the Hilda population and 9 spectra of 6 in the Cybele.
				
		The slope error for objects in our sample is described in section \ref{sec:nearirobs}. In this case all the errors are assumed to be 1\%/1000A plus the Monte-Carlo produced error, applying the same methodology as in section \ref{subsubsec:slope}. For objects with more than one observation the slopes are averaged and the error is chosen between the standard deviation or the propagated error, whichever is higher.


\section{Results}
\label{sec:results}

	\begin{figure*}[!ht]
			\centering
			\includegraphics[angle=0,scale=.55]{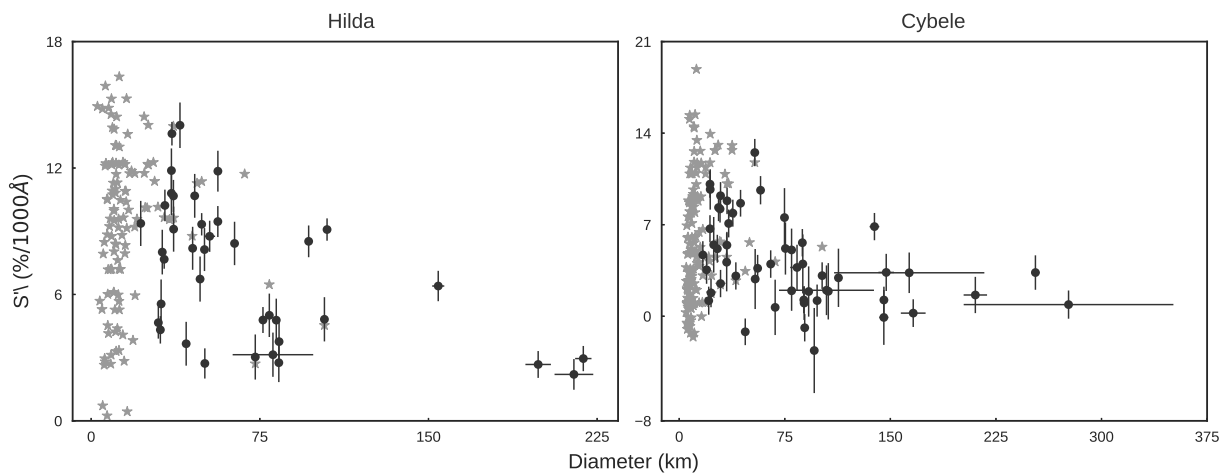} 
			\caption{Visible spectral slope $versus$ diameter for Hildas (\textit{left}) and Cybele (\textit{right}) populations. Grey stars are SDSS slopes while black points are spectral slopes.}
			\label{fig:diam}
	\end{figure*}

	The results for the analysis of the visible spectra are shown in Table \ref{table:vissoar} for Cybele and Hilda objects of our sample and in table \ref{table:viscybele}, for the Cybele and in table \ref{table:vishilda} for the  Hilda objects in the literature. Features that could not be measured due to the spectral interval are marked with a star ('*') symbol, and where it could be measured, but the feature was not detected with a dash ('-') symbol. 
	
	For an extended analysis, we add information optical geometric albedo and diameters, obtained from the current release of the NEOWISE dataset \citep{2016PDSS..247.....M}.  Tables \ref{tab:cybprop} and \ref{tab:hildaprop} list proper elements \citep{2015PDSS..234.....N} and geometric albedo for the objects in our Cybele and Hilda samples, respectively. 
	
	Figure \ref{fig:diam} shows the scatter plots of spectral slope versus diameter for the Cybele and Hilda objects. The enlarged samples show a trend that had been previously noted by other authors \citet{lagerkvist2005s} and \citet{dahlgren1997s}, in which the larger objects in those populations tend to present  intermediate values for the spectral slope, and that  the scatter in that parameter increases for smaller diameters.
	
	\begin{figure*}[!ht]
		\centering
		\includegraphics[angle=0,scale=.8]{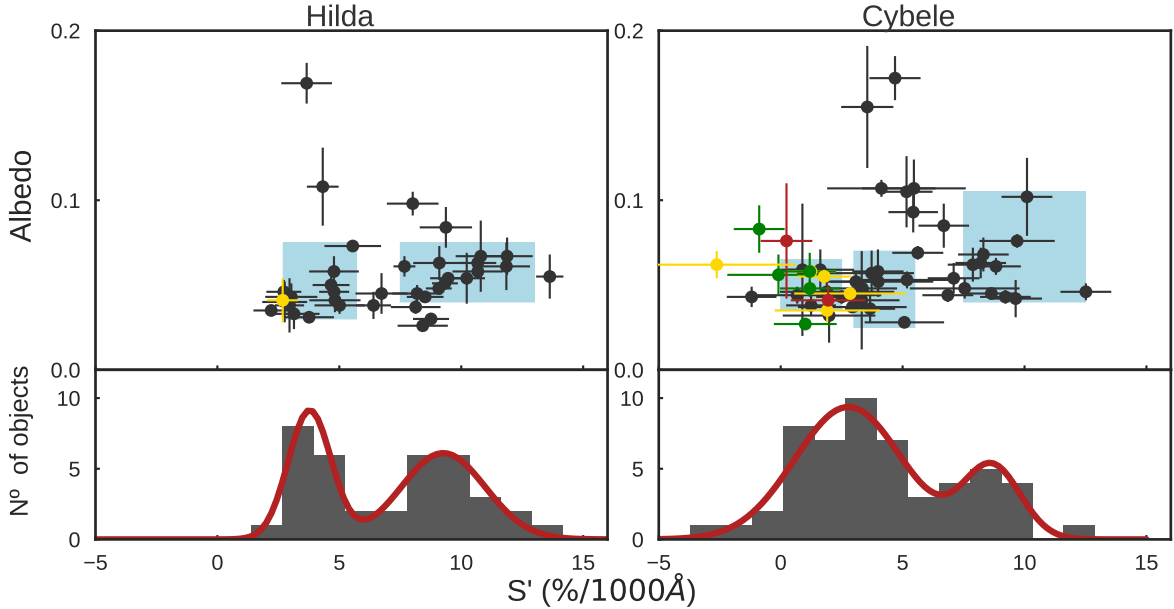} 
		\caption{\textit{Top}: Albedo-Slope distribution of Hilda and Cybele asteroids. The blue squares represents the clusters as described in section~\ref{sec:sdss}, using the SDSS and WISE data.  Yellow points are objects with UV-drop, green are objects with 0.7 $\mu$m absorption band, and red are objects that present both features. \textit{Bottom}: Histogram for visible slope distribution in the dynamical groups. Red lines represent Gaussian fitted profiles for the bimodal distributions. }
		\label{fig:slopes}			
	\end{figure*}
		
	The bottom panels of Figure \ref{fig:slopes} shows the histograms of spectral slope for the Cybele and Hilda asteroids.  On both populations it is possible to detect a bimodality on the spectral slope distribution, which had also been pointed out in the literature \citep{gil2008surface,gil2010tax}. On the top panels of this figure we show scatter plots of geometric albedo versus spectral slope.  It is possible to assure the presence of two clusters: the first, centered at $\sim4.2 \%/1000$\AA, consists of X-class asteroids, while the redder group, centered at $\sim9.4 \%/1000$\AA, is dominated by the D-class. For the Cybele group, the first cluster is centered at $\sim2.3 \%/1000$\AA, with a mix of the C-class and X-class objects, with a clear separation for the redder D-class, centered at $\sim8.8 \%/1000A$. On both groups the clusters with higher spectral slopes tend also to present slightly higher albedo than the clusters with lower spectral slope. \cite{bauer2013} and \cite{duffard2014} observed a similar behavior in the Centaur population, which also present a bimodal color distribution. Although, it is worthy to notice the redder group in the Centaur population is substantially redder than a typical D-type asteroid.   
	
	The Cybele group shows a wider variety of colors and taxonomic classes, but predominantly primitive classes (see Tables \ref{table:viscybele} and \ref{table:vishilda} ). We stress the presence of two S-type objects: (679) Hippodamia with a diameter  $D=42$ km and (3675) Kemstach with $D=18$ km, according to NEOWISE data. \citet{gil2010tax} showed other five potentially S or Q-types with SDSS data, although these are smaller objects.

	It is also important to note the distribution of objects with signs of hydration in the Cybele and Hilda groups. The 0.7 $\mu m$ absorption band is only detected in seven of the 55 Cybele asteroids, and none is observed within the Hildas. All but one are C-class objects (940 is a Xc). The UV-drop is detected in six Cybele objects, and only one Hilda: (334) Chicago. There are two Cybele objects that present both features, (121) Hermione and (168) Sibylla. The asteroids (334) and (121) are the only objects in the Hilda and the Cybele group, respectively, in \cite{takir2012outer} to show a "sharp" shape for the 3 $\mu$m feature, which is also associated to the presence of hydrated minerals. For objects that present the turn-off, we also measured the UV-slope. Since only a small amount of objects presents this feature, no relevant information was found for the UV-slope.
	
	\begin{figure}[!hb]
		\centering
		\includegraphics[angle=0,scale=.7]{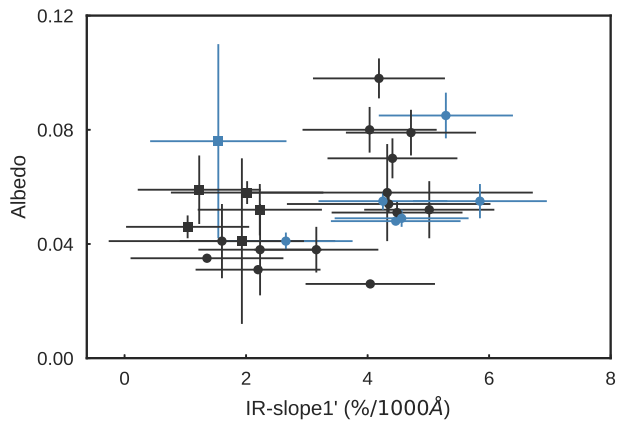} 
		\caption{Albedo $versus$ IR-slope1 for Hildas (dots) and Cybele (square) objects. Asteroids that show a decay towards redder wavelengths are labeled in blue.}
		\label{fig:ir}
	\end{figure}

	The results for the near-IR analysis for TNG spectra are presented in Table \ref{table:hildasir}, and in Table \ref{table:liteir} for the literature Hilda and Cybele objects. The majority of objects in the Hilda group present a reddish IR-slope1 and are classified as D-type objects. No D-type is observed in the Cybele group (consisting of six objects), however, this is somehow to be expected. According to \cite{demeo2013}, the dominant class for objects larger than 100 km is the P-type (equivalent to the X-type for our purpose), with very small contribution from D-type. In our sample, five out of six objects are larger than 150 km, therefore the absence of this class in our sample is consistent with the previous results.
	Figure \ref{fig:ir} suggests that the clusters in the near-infrared slope reflect in the albedo distribution, redder objects tends to higher albedo, accordantly to the behavior in figure \ref{fig:slopes}a. Objects with a significant difference from IR-slope1 to the IR-slope2 (higher than 1.5 $\%/1000$\AA) are observed in both groups.

\section{Extended analysis with large public datasets}
\label{sec:sdss}
	\begin{figure*}[!ht]
	\centering
	\includegraphics[angle=0,scale=.55]{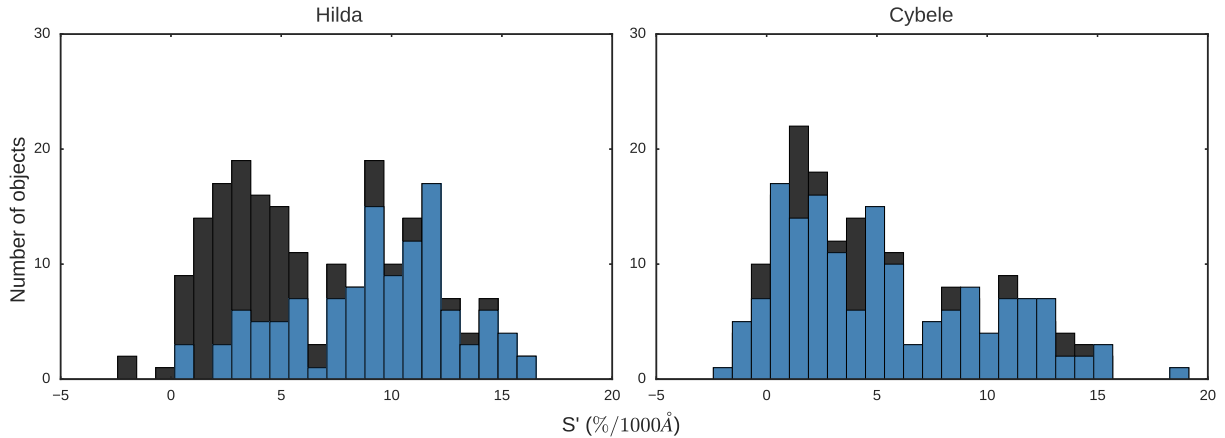} 
	\caption{Histogram for Hilda and Cybele populations with SDSS data. Blue histogram show groups with families removed}
	\label{fig:Hist}
\end{figure*}

	In this section we use data from large public databases in order to interpret our spectra within the broader context of the distribution of physical properties of the objects in the Cybele and Hilda regions, and then compare them with the physical properties of the objects in the inner edge of the 2:1 resonance, and also with the Trojan population. For spectral slope and taxonomic classification we use data from the Sloan Digital Sky Survey Moving Object Catalog \citep{2010PDSS..124.....I}  that were classified by \cite{2010A&A...510A..43C} and \cite{2011PDSS..145.....H}  into a taxonomic scheme designed to be compatible with the Bus classification, within the limits imposed by the spectral resolution of the SDSS data. Optical geometric albedo and diameter were obtained from the current release of the NEOWISE dataset \citep{2016PDSS..247.....M}, and lists of members of the dynamic asteroid families in the regions were taken from \cite{2015PDSS..234.....N}. 

	To calculate the spectral slope from the asteroid reflectance spectra listed in \cite{2011PDSS..145.....H} in a way that is compatible with the procedure described in section 3.2.2 we made a linear fit to the reflectances in the $g$, $r$, and $i$ SDSS filters (centered at 0.47, 0.62 and 0.76 $\mu$m), normalized to $g$. To calculate the slope uncertainties we created 1000 clones of each observation by drawing random values for the reflectance in each filter using normal distributions with means equal to the listed reflectance value and variances equal to the listed uncertainties. The resultant spectral slope distribution was then fitted with a Gaussian curve, whose mean and variance were then adopted as the final value for the spectral slope and its uncertainty, respectively, expressed in units of $\%/1000$\AA.  

	Using $3.3<a<3.7$ au and $3.7<a<4.5$ au to define Cybele and Hilda groups, we obtained a total of 255 asteroids listed in \cite{2015PDSS..234.....N} in the Cybele and 297 in the Hilda region with SDSS observations. Of these, 179 objects in the Cybele and 208 in the Hilda region also had tabulated albedos and diameters from \cite{2016PDSS..247.....M}. \cite{2015PDSS..234.....N} defines two families in the Cybele region, Sylvia and Ulla, and two in the Hilda region, Hilda and Schubart. In the Cybele families there were 20 objects from Sylvia and 2 from Ulla with both SDSS data and NEOWISE albedo. Similarly, 58 from Hilda and 31 from Schubart, in the Hilda region. Rejecting the object with indication of olivine/pyroxene absorption bands (members of the S complex), we are left with 177 objects in the Cybele region with slope and albedo, and 118 objects in the Hilda region.

	In order to compare with the inner and outer edge of the Cybele and Hilda populations, we also consider the slope and albedo distributions for the Trojans and for the members of the Themis family. The later is taken here as representative of the material of the inner border of the 2:1  mean motion resonance with Jupiter. We consider 575 and 330 objects with both albedo and slope from Themis dynamical family and Trojans, respectively.

	Figure \ref{fig:Hist} shows the slope distribution of the featureless asteroids on Cybele and Hilda populations, with and without family members. The distribution of slopes in these regions are clearly bimodal, as discussed in the previous section. The Hilda and Schubart dynamical families contributed strongly to the peak at lower spectral slopes, but an excess of objects with small spectral slopes remain even after the removal of the listed family members. On the other hand, the removal of nominal family members does not affect significantly the slope distribution in the Cybele region.  In what follows thus we will remove members of the families from the Hilda region, but we will keep the family members in the Cybele region. 
	
	\begin{figure*}[!ht]
	\centering
	\begin{tabular}{cc}
		{\includegraphics[angle=0,scale=.45]{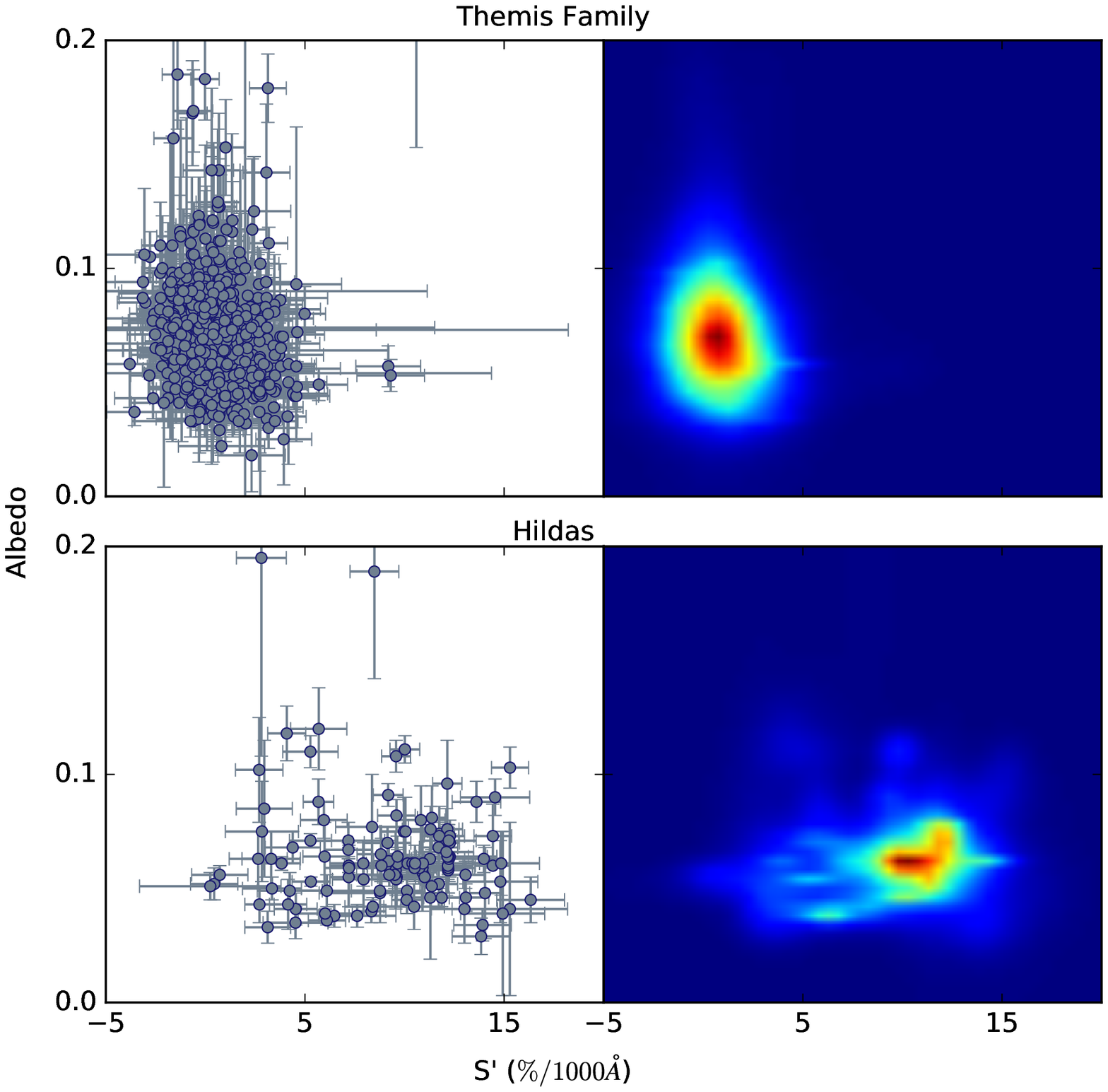}}
		{\includegraphics[angle=0,scale=.45]{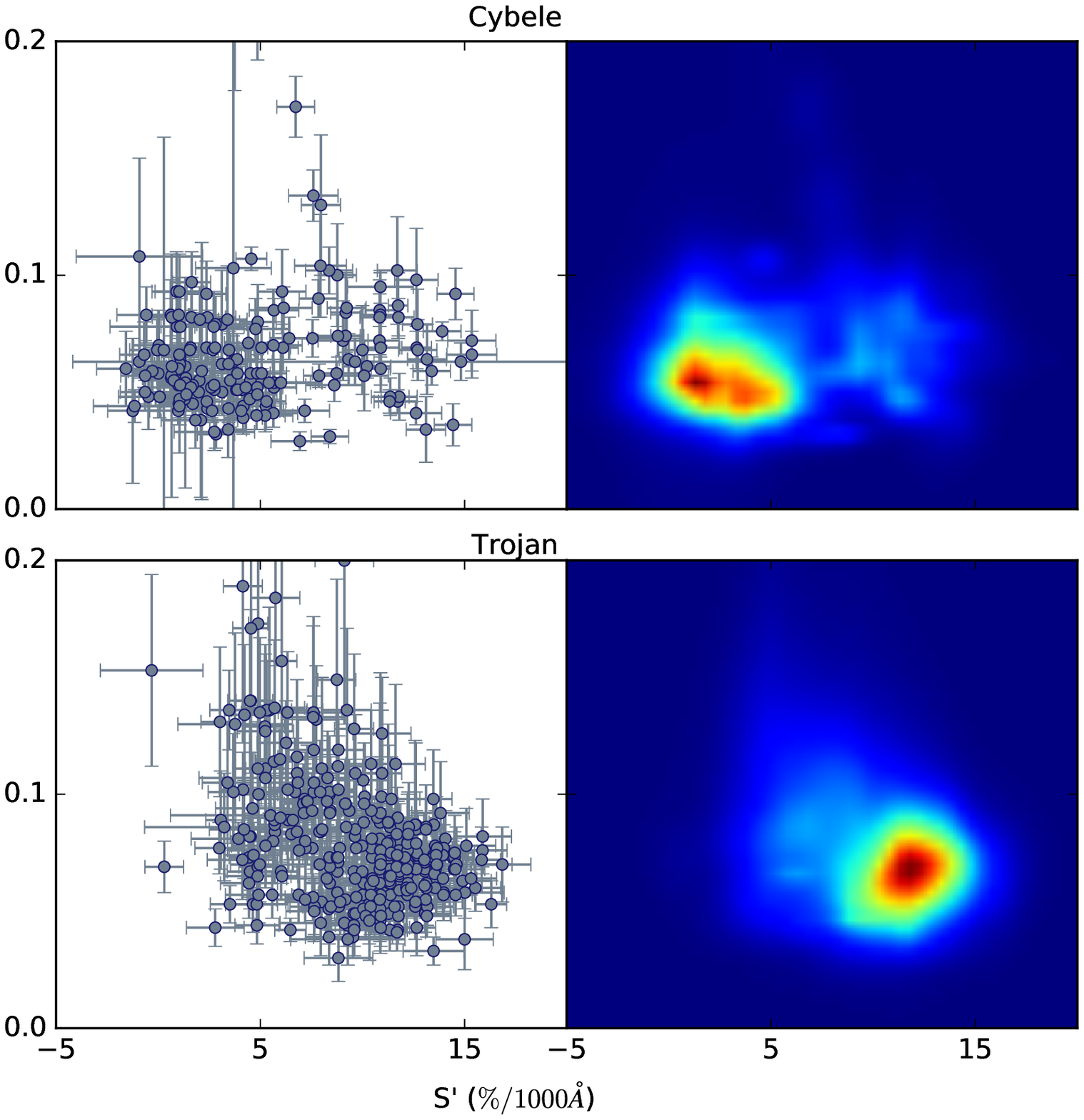} }
		
	\end{tabular}
	
	\caption{Scatter and density plots for Themis collisional family, Cybele, Hilda and Trojan populations }	
	\label{fig:density}
\end{figure*}

	These characteristics on the slope and albedo distribution on both populations had already been reported and discussed \citep{2017AJ....153...69W,2012AJ....143..141K,2012ApJ...744..197G,2011AJ....141..186R,gil2010tax,gil2008surface}, but no extensive analysis of the joined distributions of these two compositional indicators had been performed in the literature so far. To do so, we use scatter plots of slope {\it versus} albedo to construct weighted density plots. This is done  by considering that each measurement of both albedo and slope defines a gaussian function with a mean value equal to the measured value and variance equal to its uncertainty, with the total density at each possible value of slope and albedo given by the sum of the gaussians of all measurements. Peaks on these density plots correspond thus to the probability of finding members of each population on given points of the slope-albedo space. Figure \ref{fig:density} shows the scatter and density plots for all populations. A number of clusters can be seen on the density plots. In the Cybele region (Fig. \ref{fig:density}) the bimodality is apparent in the albedo-slope space. The lower slope peak clearly defined against the lower density cluster at higher slopes. The cluster at lower slopes concentrated a slightly lower albedos than the higher slope clusters. Also, the lower slope peak appears to consist of two subclusters that were not distinguishable from the spectral analysis (Section \ref{sec:results}, Fig. \ref{fig:slopes}), the most dense concentrated at lower spectral slopes and sligthtly higher albedos than the less dense one.  In the Hilda region the slope bimodality is also apparent on the density plots, with the lower spectral slope population also appearing at slightly lower albedos than the higher slope population that dominates the region, though the removal of the objects that belong to a collisional family vastly diminish the population of objects in the first cluster. Again, in the Trojan region (Figure \ref{fig:density}) the bimodality on slopes also seem to correlate with albedo, but with the lower slope group having higher albedos than the dominant higher slope group. Finally, the Themis family plots as a dense single cluster in the slope-albedo space. 

	\begin{figure}[!ht]
		\begin{center}
			\includegraphics[angle=0,scale=.7]{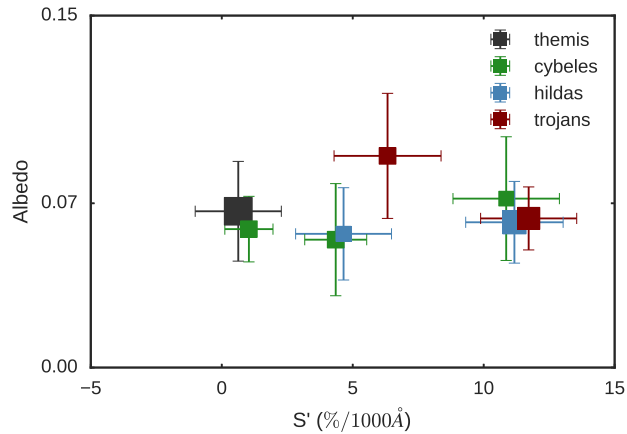} 
		\end{center}
		\vspace{-0.5cm}
		\caption{Weighted averages and standard deviation for $p_v$ and $slope$
			of objects in the Themis family and on the Hilda, Cybele and
			Trojan region. Sizes on scatter plots are related to proportion of member in each cluster, for each population.}
		\label{fig:groups}
	\end{figure}

	We then proceed to define limits for each cluster. This is done visually, using the density plots (Fig. \ref{fig:density}). For Cybele and Hilda populations this can be done simply by defining limits in slope for each cluster. The three main groups in the Cybeles are thus defined with $S'<2.5$, $2.5<S'<7.5$$\%/1000$\AA~, and $S'>7.5$$\%/1000$\AA. The two clusters in Hilda group can also be defined using $7.5$$\%/1000$\AA~ as limiting value. The Trojans, on the other hand, are better separated by a straight line given by $p_v=0.055-0.1575(S'-7)$. We can then calculate the weighted average for $p_v$ and $slope$ for each cluster and the corresponding standard deviations (Table \ref{table:clusters}). Figure \ref{fig:groups} shows the distribution of these clusters. Given that those clusters are defined mostly by ranges in the spectral gradient, they can be associated loosely with the equivalent Tholen's classes, in order to facilitate the qualitative comparison with previous works. Therefore we will be referring to the lower slope clusters as "C-cluster", to the  the intermediate slope cluster as "P-cluster" and the higher slope clusters as "D-cluster". In Figure \ref{fig:slopes} we show the specific clusters of the Cybele and Hilda groups (blue boxes) for a comparison with the spectroscopic data analysis. Although these clusters are defined visually we can see they correlate to the boxes defined in figure 5 of \cite{demeo2013} for the C, X and D types. 

			\begin{table}[!hb]
				\setcounter{table}{9}
				\centering
				\caption{Weighted averages and standard deviation for $p_v$ and $slope$ of clusters in the Themis family and on the Hilda, Cybele and Trojan region.}
				\small
				\begin{tabular}{llcccc} \hline \hline
					Population & Cluster & $\rho_g$ & $\rho_{g(std)}$ &  $S'$ &  $S'_{std}$ \\
					&  &  &  &   &    \\
					\hline
					Themis & C  & 0.066 & 0.021 & 0.63 & 1.64  \\
					Cybele & C  & 0.059 & 0.014 & 1.03 & 0.92  \\
					Cybele & P  & 0.055 & 0.023 & 4.35 & 1.18 \\
					Cybele & D  & 0.072 & 0.026 & 10.86 & 2.03  \\
					Hilda  & P   & 0.057 & 0.020 & 4.65 & 1.83  \\
					Hilda  & D   & 0.062 & 0.017 & 11.17 & 1.86  \\
					Trojan & PD & 0.090 & 0.027 & 6.32 & 2.04  \\
					Trojan & D  & 0.064 & 0.013 & 11.71 & 1.83  \\
					\hline
					
				\end{tabular}
				\label{table:clusters}
			\end{table}


	\section{Discussion}
	\label{sec:discussion}

	The joint analysis of the spectroscopic data, which represents larger objects, and spectrophotometric data, for small objects, reveals the diversity of surfaces in each group. We found two clusters in the albedo versus slope space of the Hilda group and three in the Cybele group. Although the subdivision of the lower slope cluster is only apparent using the larger SDSS sample. The bimodality is also observed in near-infrared properties. Each of this clusters might be related to different sets of compositions or processes that alter the surface of the asteroids, such as resurfacing and space weathering. 
	
	The larger variety of taxonomic classes, colors and albedo distribution reflects a wider range of possible compositions for the Cybele asteroids than Hilda asteroids. The presence of high-albedo objects ($\rho_g>0.1$) and even S-type asteroids among them suggest a contribution of objects formed in closer to Sun than their current positions.
	
	The presence of two high-albedo objects among the Hilda also suggests that there is some contribution from the main belt objects to the population, as proposed by \cite{2012ApJ...744..197G}. We investigated if the high albedos of these two objects can be explained by possible biases in their radiometric diameters and/or H magnitudes. (1162) Larissa has two groups of thermal IR observations by WISE \citep{Mainzer2011} and also by AKARI (see \cite{usui2011}), so we recomputed diameters using the NEATM implementation of \cite{victor}. Our fits to all four groups of observations are very similar to those reported by NEOWISE, and the albedos remain higher than 0.11 even if we increase the value of the H magnitude by 0.3 mag. We also reproduce the reported size and albedo for (3843) OISCA, but in this case we can reach a reasonably lower albedo of 0.08 $\pm$ 0.02 if we increase the H-value by 0.3 mag (in fact,\cite{veres2015} obtained H$=10.9 \pm 0.3$), so the high-$p_v$ value is less robust for this object.  

	The hydration band in the visible spectra has been shown common in primitive objects of other regions of the main asteroidal belt. \citet{fornasier2014hydration} finds that 45\% of C-complex asteroids presents the band in contrast with 4.5\% for P-type in the Tholen taxonomy. For the Cybele group, we found the absorption band in only $\sim30$\% of the C-types. The asteroid (940) Kordula was classified both in the Cgh and in Xc, based on the spectra obtained in this work and in \cite{vilas2006pds}, respectively. Figure \ref{fig:slopes} shows that these objects are mostly related to the C-cluster in Cybele group, and the object (334) Chicago is on the left edge of the P-cluster in the Hilda group.
	
	The small amount of hydrated objects among the Cybele asteroids and single presence of the object (334) Chicago, which presents a turn-off point in the visible, and a "sharp" 3.0 $\mu$m band \citep{takir2012outer}, the only hydrated asteroid found in the Hilda group, points to a scenario where hydration did not act strongly in these groups . In \cite{fornasier2014hydration}, the authors  analyze a dataset of over 600 spectra of primitive asteroids in the literature and conclude that the aqueous alteration process is dominant in the $2.1 - 3.1$ AU range, at smaller heliocentric distances than proposed by \cite{vilas1994iron}. \cite{morate} and Morate et al. (in prep.) shows families in the inner belt with a high amount of hydrated objects. Conjointly, thermal modeling by \cite{1993Sci...259..653G} proposes the hydration process to act in the $2.5 - 3.3$ AU range, just before the Cybele region. Thus, the Cybele group may delimit the region where aqueous alteration process could  have occurred. However, it seems that the process acts predominantly at smaller heliocentric distances than those at where the Cybele and Hilda groups are located, and it can not be ruled out that the few hydrated objects found might also be contribution of objects originated in the main-belt.

	We confirm the trend for larger objects in both groups presenting a more neutral color. In order to explain this scenario, \citet{lagerkvist2005s} and \citet{dahlgren1997s} proposed that D-types objects could be more fragile than P-types, and therefore, they can be more affected by disruptive events, and would be more numerous as smaller objects. In \citet{gil2008surface} and \citet{gil2010tax}, the authors emphasized that the trend is not seen in the small objects, observed with SDSS. They argue in favor of a combination of space weathering and resurfacing effects as the main explanation for this phenomenon.  Ion-irradiation experiments on samples of Tagish-Lake meteorite tends to neutralize the spectral slope in the visible and near infrared spectral ranges \citep{vernazza2013targish, lantz}. If the surfaces of D-type asteroids are optically dominated by similar materials, i.e., dark red hydrocarbon minerals, space weathering effects favor the evolution to a P-type surface. On the other hand, collisional disruption or collisional resurfacing would expose unweathered D-type material. Therefore they argue that the larger bodies, which did not experience catastrophic disruption or significant collisional resurfacing have more neutral colors, since their surfaces have been exposed to ion flux for longer times. On the contrary, the observed smaller objects could be fragments of larger asteroids recently disrupted by catastrophic collisions, showing fresh and more red surfaces, or have more neutral colors due to the combination of the effect produced by the ion bombardment and lack of small projectiles in the population to disrupt or resurface it, producing a color diversity in the observable small end of the size distribution.
	
	Though, \cite{vernazza2013targish} and \cite{lantz} also stated that the space weathering has a brightening effect on Tagish Lake samples. In opposition, we note that in visible and near-infrared spectra, D-cluster objects presents slightly higher albedo than less-red objects. \cite{carvano2003s} analyzed a sample of 460 featureless spectra asteroids from all regions of the main belt and found a similar behavior. The ambient effects, such as space weathering and collisional resurfacing should diversify more the spectral gradient in the smaller objects than in the larger ones for the reasons pointed in \cite{gil2008surface} and \cite{gil2010tax}, but the same trend can also be explained if there is more than one compositional group in the population.  
	
	The red color objects are commonly hypothesized to be similar in composition to the Tagish Lake meteorite, that presents a red spectrum and a very low albedo.	On \cite{takir2012outer}, the authors argue that all the observed D- and P-types located in the $3.0<a<4.0$ au region exhibit a rounded shape 3.0 $\mu$m. They give a possible explanation for the feature with a thin layer of water frost in the surface of these asteroids. The presence of ice in the surface of the objects could also explain the higher albedo of D-types objects than Tagish Lake meteorite, though it is not clear how the presence of frost in the asteroid surface should alter the spectral slope. 
				
	To compare these groups to the neighboring populations, we analyzed the Themis and Trojan colors and albedo distribution. Figure \ref{fig:groups} shows the measured center for each cluster in the Themis family, and in the Cybele, Hilda and Trojans populations. The values can be seen in Table \ref{table:clusters}. There is a clear match for the three clusters in the Cybele group, which presents objects similar to those of Themis family and both groups observed in the Hildas. In the Jupiter Trojans it is possible to identify that the D-cluster matches objects with a cluster in Cybele and Hilda, but there is one group of objects which is not statistically strong in any of previous populations. We shall call it DP-cluster. Trojans DP-cluster seems to be redder and present higher albedo than the P-cluster in the Hilda and Cybele groups. A Two-dimensional Kolmogorow-Smirnov test rejects the hypothesis that the DP-cluster of the Trojan population comes from a similar distribution of the P-cluster of the Hilda os Cybele populations, providing a p-value $<<0.01$.   
	
	Planetary migration models suggests a common origin to Hilda and Trojans groups. Though, despite the fact that both populations presents a bimodal distribution, as also discussed in \cite{2017AJ....153...69W},  in the slope versus albedo space, they do not seem to be matching groups. A possible explanation is that Trojans asteroids may suffer resurfacing more frequently than Hilda asteroids, and present a generally younger surface. \cite{2002aste.book..545D} argues that the current intrinsic collisional probability and impact velocities are significantly lower for the Hildas than for the Trojans. Though, one of the Trojans groups is in good agreement with one of the Hilda clusters. Another possible explanation is that Hilda and Trojans have objects of different compositions and origins. The apparent continuity of asteroids surfaces and density objects from the Themis family to the Cybele, Hilda and Trojan populations may suggest a gradient of composition. This scenario would impose an obstacle for planetary migration models.

	\section{Conclusions}
	\label{sec:conclusions}
		
	We obtained 18 visible and 22 near-infrared spectra of Cybele and Hilda populations at the outer edge of asteroid belt, in order to study their surface properties distributions. The sample was enlarged with literature spectra, resulting in a total of 85 visible and nine near-infrared spectra for Cybele group, and 83 visible and 35 near-infrared spectra for Hilda group. The analysis was enhanced with NEOWISE and SDSS data, for information on the optical geometrical albedo and spectrophotometric properties of the small size objects in these populations. We conclude that:
	\begin{itemize}

	\item The Hilda population shows a bimodal distribution of surface properties, while in the Cybele we could identify three predominant groups of objects. The Cybele population shows a wider contribution of neutral color objects than the Hilda. The bimodality is also observed in the near-infrared analysis, where we observe a trend of redder objects showing higher albedo.
	
	\item The Cybele group presents only 9 out 55 asteroids with evidence of hydrated minerals on their surfaces, while in the Hilda group only in the object (334) Chicago the presence of aqueous altered minerals in the surface can be confirmed. Therefore, the Cybele population could possibly delimit the outer edge where the aqueous alteration process can act strongly.
	
	\item We identify a continuity of surface properties from the Themis family to the Cybele, Hilda and Trojan populations. The last two populations shows distinct distribution of surface properties. This result could be related to a compositional gradient. 
	
	\end{itemize}

	\bigskip
	\textbf{Acknowledgments:}
	 Based on observations obtained at the Southern Astrophysical Research (SOAR) telescope, which is a joint project of the Minist\'erio da Ci\^encia, Tecnologia, e Inovação (MCTI) da Rep\'ublica Federativa do Brasil, the U.S. National Optical Astronomy Observatory (NOAO), the University of North Carolina at Chapel Hill (UNC), and Michigan State University.
	
	 M. N. De Pr\'a acknowledges support from the CAPES (Brazil).
	 
	 J. M. Carvano acknowledges support from the CNPq (Brazil).
	
	 J. Licandro, and J. de Le\'on acknowledge support from the AYA2015-67772-R (MINECO, Spain).
	 
	 V. Al\'i-Lagoa: The research leading to these results has received funding from the European Union’s Horizon 2020 Research and Innovation Programme, under Grant Agreement $n^o$ 687378.
	\bibliographystyle{elsarticle-harv} 
	\bibliography{ref}

\begin{thebibliography}{63}
\expandafter\ifx\csname natexlab\endcsname\relax\def\natexlab#1{#1}\fi
\expandafter\ifx\csname url\endcsname\relax
  \def\url#1{\texttt{#1}}\fi
\expandafter\ifx\csname urlprefix\endcsname\relax\def\urlprefix{URL }\fi

\bibitem[{{Al{\'{\i}}-Lagoa} et~al.(2016){Al{\'{\i}}-Lagoa}, {Licandro},
  {Gil-Hutton}, {Ca{\~n}ada-Assandri}, {Delbo'}, {de Le{\'o}n}, {Campins},
  {Pinilla-Alonso}, {Kelley}, and {Hanu{\v s}}}]{victor}
{Al{\'{\i}}-Lagoa}, V., {Licandro}, J., {Gil-Hutton}, R.,
  {Ca{\~n}ada-Assandri}, M., {Delbo'}, M., {de Le{\'o}n}, J., {Campins}, H.,
  {Pinilla-Alonso}, N., {Kelley}, M.~S.~P., {Hanu{\v s}}, J., Jun. 2016.
  {Differences between the Pallas collisional family and similarly sized B-type
  asteroids}. \aap 591, A14.

\bibitem[{{Bauer} et~al.(2013){Bauer}, {Grav}, {Blauvelt}, {Mainzer},
  {Masiero}, {Stevenson}, {Kramer}, {Fern{\'a}ndez}, {Lisse}, {Cutri},
  {Weissman}, {Dailey}, {Masci}, {Walker}, {Waszczak}, {Nugent}, {Meech},
  {Lucas}, {Pearman}, {Wilkins}, {Watkins}, {Kulkarni}, {Wright}, {WISE Team},
  and {PTF Team}}]{bauer2013}
{Bauer}, J.~M., {Grav}, T., {Blauvelt}, E., {Mainzer}, A.~K., {Masiero}, J.~R.,
  {Stevenson}, R., {Kramer}, E., {Fern{\'a}ndez}, Y.~R., {Lisse}, C.~M.,
  {Cutri}, R.~M., {Weissman}, P.~R., {Dailey}, J.~W., {Masci}, F.~J., {Walker},
  R., {Waszczak}, A., {Nugent}, C.~R., {Meech}, K.~J., {Lucas}, A., {Pearman},
  G., {Wilkins}, A., {Watkins}, J., {Kulkarni}, S., {Wright}, E.~L., {WISE
  Team}, {PTF Team}, Aug. 2013. {Centaurs and Scattered Disk Objects in the
  Thermal Infrared: Analysis of WISE/NEOWISE Observations}. \apj 773, 22.

\bibitem[{Bell(1989)}]{bell1989m}
Bell, J.~F., 1989. Mineralogical clues to the origins of asteroid dynamical
  families. Icarus 78~(2), 426--440.

\bibitem[{Bro{\v{z}} and Vokrouhlick{\`y}(2008)}]{brovz2008asteroid}
Bro{\v{z}}, M., Vokrouhlick{\`y}, D., 2008. Asteroid families in the
  first-order resonances with jupiter. Monthly Notices of the Royal
  Astronomical Society 390~(2), 715--732.

\bibitem[{{Bus} and {Binzel}(2004)}]{smassii}
{Bus}, S., {Binzel}, R.~P., Oct. 2004. {Small Main-belt Asteroid Spectroscopic
  Survey, Phase II}. NASA Planetary Data System 1.

\bibitem[{{Bus} and {Binzel}(2002)}]{bus2002tax}
{Bus}, S.~J., {Binzel}, R.~P., Jul. 2002. {Phase II of the Small Main-Belt
  Asteroid Spectroscopic Survey. A Feature-Based Taxonomy}. \icarus 158,
  146--177.

\bibitem[{{Campins} et~al.(2010){Campins}, {Hargrove}, {Pinilla-Alonso},
  {Howell}, {Kelley}, {Licandro}, {Moth{\'e}-Diniz}, {Fern{\'a}ndez}, and
  {Ziffer}}]{campins2010themis}
{Campins}, H., {Hargrove}, K., {Pinilla-Alonso}, N., {Howell}, E.~S., {Kelley},
  M.~S., {Licandro}, J., {Moth{\'e}-Diniz}, T., {Fern{\'a}ndez}, Y., {Ziffer},
  J., Apr. 2010. {Water ice and organics on the surface of the asteroid 24
  Themis}. \nat 464, 1320--1321.

\bibitem[{Carvano et~al.(2003)Carvano, Moth{\'e}-Diniz, and
  Lazzaro}]{carvano2003s}
Carvano, J., Moth{\'e}-Diniz, T., Lazzaro, D., 2003. Search for relations among
  a sample of 460 asteroids with featureless spectra. Icarus 161~(2), 356--382.

\bibitem[{{Carvano} et~al.(2010){Carvano}, {Hasselmann}, {Lazzaro}, and
  {Moth{\'e}-Diniz}}]{2010A&A...510A..43C}
{Carvano}, J.~M., {Hasselmann}, P.~H., {Lazzaro}, D., {Moth{\'e}-Diniz}, T.,
  Feb. 2010. {SDSS-based taxonomic classification and orbital distribution of
  main belt asteroids}. \aap 510, A43.

\bibitem[{{Cloutis} et~al.(2011{\natexlab{a}}){Cloutis}, {Hiroi}, {Gaffey},
  {Alexander}, and {Mann}}]{cloutis2011-ci}
{Cloutis}, E.~A., {Hiroi}, T., {Gaffey}, M.~J., {Alexander}, C.~M.~O.~.,
  {Mann}, P., Mar. 2011{\natexlab{a}}. {Spectral reflectance properties of
  carbonaceous chondrites: 1. CI chondrites}. \icarus 212, 180--209.

\bibitem[{{Cloutis} et~al.(2011{\natexlab{b}}){Cloutis}, {Hudon}, {Hiroi},
  {Gaffey}, and {Mann}}]{cloutis2011-cm}
{Cloutis}, E.~A., {Hudon}, P., {Hiroi}, T., {Gaffey}, M.~J., {Mann}, P., Nov.
  2011{\natexlab{b}}. {Spectral reflectance properties of carbonaceous
  chondrites: 2. CM chondrites}. \icarus 216, 309--346.

\bibitem[{Dahlgren and Lagerkvist(1995)}]{dahlgren1995study}
Dahlgren, M., Lagerkvist, C.-I., 1995. A study of hilda asteroids. i. ccd
  spectroscopy of hilda asteroids. Astronomy and Astrophysics 302, 907.

\bibitem[{Dahlgren et~al.(1997)Dahlgren, Lagerkvist, Fitzsimmons, Williams, and
  Gordon}]{dahlgren1997s}
Dahlgren, M., Lagerkvist, C.-I., Fitzsimmons, A., Williams, I., Gordon, M.,
  1997. A study of hilda asteroids. ii. compositional implications from optical
  spectroscopy. Astronomy and Astrophysics 323, 606--619.

\bibitem[{{Davis} et~al.(2002){Davis}, {Durda}, {Marzari}, {Campo Bagatin}, and
  {Gil-Hutton}}]{2002aste.book..545D}
{Davis}, D.~R., {Durda}, D.~D., {Marzari}, F., {Campo Bagatin}, A.,
  {Gil-Hutton}, R., Mar. 2002. {Collisional Evolution of Small-Body
  Populations}. pp. 545--558.

\bibitem[{{DeMeo} et~al.(2009){DeMeo}, {Binzel}, {Slivan}, and
  {Bus}}]{demeo2009tax}
{DeMeo}, F.~E., {Binzel}, R.~P., {Slivan}, S.~M., {Bus}, S.~J., Jul. 2009. {An
  extension of the Bus asteroid taxonomy into the near-infrared}. \icarus 202,
  160--180.

\bibitem[{{DeMeo} and {Carry}(2013)}]{demeo2013}
{DeMeo}, F.~E., {Carry}, B., Sep. 2013. {The taxonomic distribution of
  asteroids from multi-filter all-sky photometric surveys}. \icarus 226,
  723--741.

\bibitem[{{Di Sisto} et~al.(2005){Di Sisto}, {Brunini}, {Dirani}, and
  {Orellana}}]{disisto2005hildacomet}
{Di Sisto}, R.~P., {Brunini}, A., {Dirani}, L.~D., {Orellana}, R.~B., Mar.
  2005. {Hilda asteroids among Jupiter family comets}. \icarus 174, 81--89.

\bibitem[{{Duffard} et~al.(2014){Duffard}, {Pinilla-Alonso}, {Santos-Sanz},
  {Vilenius}, {Ortiz}, {Mueller}, {Fornasier}, {Lellouch}, {Mommert}, {Pal},
  {Kiss}, {Mueller}, {Stansberry}, {Delsanti}, {Peixinho}, and
  {Trilling}}]{duffard2014}
{Duffard}, R., {Pinilla-Alonso}, N., {Santos-Sanz}, P., {Vilenius}, E.,
  {Ortiz}, J.~L., {Mueller}, T., {Fornasier}, S., {Lellouch}, E., {Mommert},
  M., {Pal}, A., {Kiss}, C., {Mueller}, M., {Stansberry}, J., {Delsanti}, A.,
  {Peixinho}, N., {Trilling}, D., Apr. 2014. {``TNOs are Cool'': A survey of
  the trans-Neptunian region. XI. A Herschel-PACS view of 16 Centaurs}. \aap
  564, A92.

\bibitem[{Dumas et~al.(1998)Dumas, Owen, and Barucci}]{dumas1998near}
Dumas, C., Owen, T., Barucci, M., 1998. Near-infrared spectroscopy of
  low-albedo surfaces of the solar system: Search for the spectral signature of
  dark material. Icarus 133~(2), 221--232.

\bibitem[{Emery et~al.(2006)Emery, Cruikshank, and
  Van~Cleve}]{emery2006thermal}
Emery, J., Cruikshank, D., Van~Cleve, J., 2006. Thermal emission spectroscopy
  (5.2--38 $\mu$m) of three trojan asteroids with the spitzer space telescope:
  Detection of fine-grained silicates. Icarus 182~(2), 496--512.

\bibitem[{{Fornasier} et~al.(2014){Fornasier}, {Lantz}, {Barucci}, and
  {Lazzarin}}]{fornasier2014hydration}
{Fornasier}, S., {Lantz}, C., {Barucci}, M.~A., {Lazzarin}, M., May 2014.
  {Aqueous alteration on main belt primitive asteroids: Results from visible
  spectroscopy}. \icarus 233, 163--178.

\bibitem[{Gaffey et~al.(1989)Gaffey, Bell, and Cruikshank}]{gaffey1989}
Gaffey, M.~J., Bell, J.~F., Cruikshank, D.~P., 1989. Reflectance spectroscopy
  and asteroid surface mineralogy. In: Asteroids II. pp. 98--127.

\bibitem[{Gil-Hutton and Brunini(2008)}]{gil2008surface}
Gil-Hutton, R., Brunini, A., 2008. Surface composition of hilda asteroids from
  the analysis of the sloan digital sky survey colors. Icarus 193~(2),
  567--571.

\bibitem[{Gil-Hutton and Licandro(2010)}]{gil2010tax}
Gil-Hutton, R., Licandro, J., 2010. Taxonomy of asteroids in the cybele region
  from the analysis of the sloan digital sky survey colors. Icarus 206~(2),
  729--734.

\bibitem[{Gomes et~al.(2005)Gomes, Levison, Tsiganis, and
  Morbidelli}]{gomes2005origin}
Gomes, R., Levison, H.~F., Tsiganis, K., Morbidelli, A., 2005. Origin of the
  cataclysmic late heavy bombardment period of the terrestrial planets. Nature
  435~(7041), 466--469.

\bibitem[{{Grav} et~al.(2012){Grav}, {Mainzer}, {Bauer}, {Masiero}, {Spahr},
  {McMillan}, {Walker}, {Cutri}, {Wright}, {Eisenhardt}, {Blauvelt}, {DeBaun},
  {Elsbury}, {Gautier}, {Gomillion}, {Hand}, and
  {Wilkins}}]{2012ApJ...744..197G}
{Grav}, T., {Mainzer}, A.~K., {Bauer}, J., {Masiero}, J., {Spahr}, T.,
  {McMillan}, R.~S., {Walker}, R., {Cutri}, R., {Wright}, E., {Eisenhardt},
  P.~R., {Blauvelt}, E., {DeBaun}, E., {Elsbury}, D., {Gautier}, T.,
  {Gomillion}, S., {Hand}, E., {Wilkins}, A., Jan. 2012. {WISE/NEOWISE
  Observations of the Hilda Population: Preliminary Results}. \apj 744, 197.

\bibitem[{{Grimm} and {McSween}(1993)}]{1993Sci...259..653G}
{Grimm}, R.~E., {McSween}, H.~Y., Jan. 1993. {Heliocentric zoning of the
  asteroid belt by aluminum-26 heating}. Science 259, 653--655.

\bibitem[{Hargrove et~al.(2012)Hargrove, Kelley, Campins, Licandro, and
  Emery}]{hargrove2012asteroids}
Hargrove, K.~D., Kelley, M.~S., Campins, H., Licandro, J., Emery, J., 2012.
  Asteroids (65) cybele,(107) camilla and (121) hermione: Infrared spectral
  diversity among the cybeles. Icarus 221~(1), 453--455.

\bibitem[{{Hasselmann} et~al.(2011){Hasselmann}, {Carvano}, and
  {Lazzaro}}]{2011PDSS..145.....H}
{Hasselmann}, P.~H., {Carvano}, J.~M., {Lazzaro}, D., Jun. 2011. {SDSS-based
  Asteroid Taxonomy V1.0}. NASA Planetary Data System 145.

\bibitem[{Hiroi et~al.(2001)Hiroi, Zolensky, and Pieters}]{hiroi2001tagish}
Hiroi, T., Zolensky, M.~E., Pieters, C.~M., 2001. The tagish lake meteorite: A
  possible sample from a d-type asteroid. Science 293~(5538), 2234--2236.

\bibitem[{{Ivezic} et~al.(2010){Ivezic}, {Juric}, {Lupton}, {Tabachnik},
  {Quinn}, and {SDSS Collaboration}}]{2010PDSS..124.....I}
{Ivezic}, Z., {Juric}, M., {Lupton}, R.~H., {Tabachnik}, S., {Quinn}, T., {SDSS
  Collaboration}, Aug. 2010. {SDSS Moving Object Catalog V3.0}. NASA Planetary
  Data System 124.

\bibitem[{{Jewitt}(2002)}]{jewitt2002slope}
{Jewitt}, D.~C., Feb. 2002. {From Kuiper Belt Object to Cometary Nucleus: The
  Missing Ultrared Matter}. \aj 123, 1039--1049.

\bibitem[{{Kasuga} et~al.(2012){Kasuga}, {Usui}, {Hasegawa}, {Kuroda},
  {Ootsubo}, {M{\"u}ller}, and {Ishiguro}}]{2012AJ....143..141K}
{Kasuga}, T., {Usui}, F., {Hasegawa}, S., {Kuroda}, D., {Ootsubo}, T.,
  {M{\"u}ller}, T.~G., {Ishiguro}, M., Jun. 2012. {AKARI/AcuA Physical Studies
  of the Cybele Asteroid Family}. \aj 143, 141.

\bibitem[{{Krot} et~al.(2015){Krot}, {Nagashima}, {Alexander}, {Ciesla},
  {Fujiya}, and {Bonal}}]{krot2015}
{Krot}, A.~N., {Nagashima}, K., {Alexander}, C.~M.~O., {Ciesla}, F.~J.,
  {Fujiya}, W., {Bonal}, L., 2015. {Sources of Water and Aqueous Activity on
  the Chondrite Parent Asteroids}. pp. 635--660.

\bibitem[{Lagerkvist et~al.(2005)Lagerkvist, Moroz, Nathues, Erikson, Lahulla,
  Karlsson, and Dahlgren}]{lagerkvist2005s}
Lagerkvist, C.-I., Moroz, L., Nathues, A., Erikson, A., Lahulla, F., Karlsson,
  O., Dahlgren, M., 2005. A study of cybele asteroids-ii. spectral properties
  of cybele asteroids. Astronomy \& Astrophysics 432~(1), 349--354.

\bibitem[{{Lantz} et~al.(2017){Lantz}, {Brunetto}, {Barucci}, {Fornasier},
  {Baklouti}, {Bour{\c c}ois}, and {Godard}}]{lantz}
{Lantz}, C., {Brunetto}, R., {Barucci}, M.~A., {Fornasier}, S., {Baklouti}, D.,
  {Bour{\c c}ois}, J., {Godard}, M., Mar. 2017. {Ion irradiation of
  carbonaceous chondrites: A new view of space weathering on primitive
  asteroids}. \icarus 285, 43--57.

\bibitem[{{Lazzaro} et~al.(2004){Lazzaro}, {Angeli}, {Carvano},
  {Moth{\'e}-Diniz}, {Duffard}, and {Florczak}}]{lazzaro2004s30s2}
{Lazzaro}, D., {Angeli}, C.~A., {Carvano}, J.~M., {Moth{\'e}-Diniz}, T.,
  {Duffard}, R., {Florczak}, M., Nov. 2004. {S $^{3}$OS $^{2}$: the visible
  spectroscopic survey of 820 asteroids}. \icarus 172, 179--220.

\bibitem[{Levison and Duncan(1993)}]{levison1993gravitational}
Levison, H., Duncan, M., 1993. The gravitational sculpting of the kuiper belt.
  The Astrophysical Journal 406, L35--L38.

\bibitem[{Licandro et~al.(2008)Licandro, Alvarez-Candal, De~Le{\'o}n,
  Pinilla-Alonso, Lazzaro, and Campins}]{licandro2008spectral}
Licandro, J., Alvarez-Candal, A., De~Le{\'o}n, J., Pinilla-Alonso, N., Lazzaro,
  D., Campins, H., 2008. Spectral properties of asteroids in cometary orbits.
  Astronomy \& Astrophysics 481~(3), 861--877.

\bibitem[{{Licandro} et~al.(2011){Licandro}, {Campins}, {Kelley}, {Hargrove},
  {Pinilla-Alonso}, {Cruikshank}, {Rivkin}, and
  {Emery}}]{licandro2011cybeleice}
{Licandro}, J., {Campins}, H., {Kelley}, M., {Hargrove}, K., {Pinilla-Alonso},
  N., {Cruikshank}, D., {Rivkin}, A.~S., {Emery}, J., Jan. 2011. {(65) Cybele:
  detection of small silicate grains, water-ice, and organics}. \aap 525, A34.

\bibitem[{{Mainzer} et~al.(2011){Mainzer}, {Bauer}, {Grav}, {Masiero}, {Cutri},
  {Dailey}, {Eisenhardt}, {McMillan}, {Wright}, {Walker}, {Jedicke}, {Spahr},
  {Tholen}, {Alles}, {Beck}, {Brandenburg}, {Conrow}, {Evans}, {Fowler},
  {Jarrett}, {Marsh}, {Masci}, {McCallon}, {Wheelock}, {Wittman}, {Wyatt},
  {DeBaun}, {Elliott}, {Elsbury}, {Gautier}, {Gomillion}, {Leisawitz},
  {Maleszewski}, {Micheli}, and {Wilkins}}]{Mainzer2011}
{Mainzer}, A., {Bauer}, J., {Grav}, T., {Masiero}, J., {Cutri}, R.~M.,
  {Dailey}, J., {Eisenhardt}, P., {McMillan}, R.~S., {Wright}, E., {Walker},
  R., {Jedicke}, R., {Spahr}, T., {Tholen}, D., {Alles}, R., {Beck}, R.,
  {Brandenburg}, H., {Conrow}, T., {Evans}, T., {Fowler}, J., {Jarrett}, T.,
  {Marsh}, K., {Masci}, F., {McCallon}, H., {Wheelock}, S., {Wittman}, M.,
  {Wyatt}, P., {DeBaun}, E., {Elliott}, G., {Elsbury}, D., {Gautier}, IV, T.,
  {Gomillion}, S., {Leisawitz}, D., {Maleszewski}, C., {Micheli}, M.,
  {Wilkins}, A., Apr. 2011. {Preliminary Results from NEOWISE: An Enhancement
  to the Wide-field Infrared Survey Explorer for Solar System Science}. \apj
  731, 53.

\bibitem[{{Mainzer} et~al.(2016){Mainzer}, {Bauer}, {Cutri}, {Grav}, {Kramer},
  {Masiero}, {Nugent}, {Sonnett}, {Stevenson}, and
  {Wright}}]{2016PDSS..247.....M}
{Mainzer}, A.~K., {Bauer}, J.~M., {Cutri}, R.~M., {Grav}, T., {Kramer}, E.~A.,
  {Masiero}, J.~R., {Nugent}, C.~R., {Sonnett}, S.~M., {Stevenson}, R.~A.,
  {Wright}, E.~L., Jun. 2016. {NEOWISE Diameters and Albedos V1.0}. NASA
  Planetary Data System 247.

\bibitem[{Mallat(1999)}]{mallat}
Mallat, S., 1999. Wavelet analysis \& its applications.

\bibitem[{{Morate} et~al.(2016){Morate}, {de Le{\'o}n}, {De Pr{\'a}},
  {Licandro}, {Cabrera-Lavers}, {Campins}, {Pinilla-Alonso}, and
  {Al{\'{\i}}-Lagoa}}]{morate}
{Morate}, D., {de Le{\'o}n}, J., {De Pr{\'a}}, M., {Licandro}, J.,
  {Cabrera-Lavers}, A., {Campins}, H., {Pinilla-Alonso}, N.,
  {Al{\'{\i}}-Lagoa}, V., Feb. 2016. {Compositional study of asteroids in the
  Erigone collisional family using visible spectroscopy at the 10.4 m GTC}.
  \aap 586, A129.

\bibitem[{Morbidelli et~al.(2005)Morbidelli, Levison, Tsiganis, and
  Gomes}]{morbidelli2005}
Morbidelli, A., Levison, H., Tsiganis, K., Gomes, R., 2005. Chaotic capture of
  jupiter's trojan asteroids in the early solar system. Nature 435~(7041),
  462--465.

\bibitem[{{Moth{\'e}-Diniz}(2010)}]{thais}
{Moth{\'e}-Diniz}, T., 2010. {Searching for minor absorptions on D-type
  asteroids}. In: {Fernandez}, J.~A., {Lazzaro}, D., {Prialnik}, D., {Schulz},
  R. (Eds.), Icy Bodies of the Solar System. Vol. 263 of IAU Symposium. pp.
  231--236.

\bibitem[{{Nesvorny}(2015)}]{2015PDSS..234.....N}
{Nesvorny}, D., Dec. 2015. {Nesvorny HCM Asteroid Families V3.0}. NASA
  Planetary Data System 234.

\bibitem[{{Popescu} et~al.(2012){Popescu}, {Birlan}, and
  {Nedelcu}}]{popescu2012mast}
{Popescu}, M., {Birlan}, M., {Nedelcu}, D.~A., Aug. 2012. {Modeling of asteroid
  spectra - M4AST}. \aap 544, A130.

\bibitem[{{Reddy} and {Sanchez}(2016)}]{reddypds}
{Reddy}, V., {Sanchez}, J.~A., Aug. 2016. {Reddy Main Belt Asteroid Spectra
  V1.0}. NASA Planetary Data System 242.

\bibitem[{{Rivkin} et~al.(2015){Rivkin}, {Campins}, {Emery}, {Howell},
  {Licandro}, {Takir}, and {Vilas}}]{rivkin2015}
{Rivkin}, A.~S., {Campins}, H., {Emery}, J.~P., {Howell}, E.~S., {Licandro},
  J., {Takir}, D., {Vilas}, F., 2015. {Astronomical Observations of Volatiles
  on Asteroids}. pp. 65--87.

\bibitem[{Rivkin and Emery(2010)}]{rivkin2010}
Rivkin, A.~S., Emery, J.~P., 2010. Detection of ice and organics on an
  asteroidal surface. Nature 464~(7293), 1322--1323.

\bibitem[{Roig and Nesvorn{\`y}(2015)}]{roig2015evolution}
Roig, F., Nesvorn{\`y}, D., 2015. The evolution of asteroids in the
  jumping-jupiter migration model. The Astronomical Journal 150~(6), 186.

\bibitem[{{Ryan} and {Woodward}(2011)}]{2011AJ....141..186R}
{Ryan}, E.~L., {Woodward}, C.~E., Jun. 2011. {Albedos of Small Hilda Group
  Asteroids as Revealed by Spitzer}. \aj 141, 186.

\bibitem[{Takir and Emery(2012)}]{takir2012outer}
Takir, D., Emery, J.~P., 2012. Outer main belt asteroids: Identification and
  distribution of four 3-$\mu$m spectral groups. Icarus 219~(2), 641--654.

\bibitem[{Tholen(1984)}]{tholen1984asteroid}
Tholen, D.~J., 1984. Asteroid taxonomy from cluster analysis of photometry.

\bibitem[{Tsiganis et~al.(2005)Tsiganis, Gomes, Morbidelli, and
  Levison}]{tsiganis2005origin}
Tsiganis, K., Gomes, R., Morbidelli, A., Levison, H., 2005. Origin of the
  orbital architecture of the giant planets of the solar system. Nature
  435~(7041), 459--461.

\bibitem[{{Usui} et~al.(2011){Usui}, {Kuroda}, {M{\"u}ller}, {Hasegawa},
  {Ishiguro}, {Ootsubo}, {Ishihara}, {Kataza}, {Takita}, {Oyabu}, {Ueno},
  {Matsuhara}, and {Onaka}}]{usui2011}
{Usui}, F., {Kuroda}, D., {M{\"u}ller}, T.~G., {Hasegawa}, S., {Ishiguro}, M.,
  {Ootsubo}, T., {Ishihara}, D., {Kataza}, H., {Takita}, S., {Oyabu}, S.,
  {Ueno}, M., {Matsuhara}, H., {Onaka}, T., Oct. 2011. {Asteroid Catalog Using
  Akari: AKARI/IRC Mid-Infrared Asteroid Survey}. \pasj 63, 1117--1138.

\bibitem[{{Vere{\v s}} et~al.(2015){Vere{\v s}}, {Jedicke}, {Fitzsimmons},
  {Denneau}, {Granvik}, {Bolin}, {Chastel}, {Wainscoat}, {Burgett}, {Chambers},
  {Flewelling}, {Kaiser}, {Magnier}, {Morgan}, {Price}, {Tonry}, and
  {Waters}}]{veres2015}
{Vere{\v s}}, P., {Jedicke}, R., {Fitzsimmons}, A., {Denneau}, L., {Granvik},
  M., {Bolin}, B., {Chastel}, S., {Wainscoat}, R.~J., {Burgett}, W.~S.,
  {Chambers}, K.~C., {Flewelling}, H., {Kaiser}, N., {Magnier}, E.~A.,
  {Morgan}, J.~S., {Price}, P.~A., {Tonry}, J.~L., {Waters}, C., Nov. 2015.
  {Absolute magnitudes and slope parameters for 250,000 asteroids observed by
  Pan-STARRS PS1 - Preliminary results}. \icarus 261, 34--47.

\bibitem[{{Vernazza} et~al.(2013){Vernazza}, {Fulvio}, {Brunetto}, {Emery},
  {Dukes}, {Cipriani}, {Witasse}, {Schaible}, {Zanda}, {Strazzulla}, and
  {Baragiola}}]{vernazza2013targish}
{Vernazza}, P., {Fulvio}, D., {Brunetto}, R., {Emery}, J.~P., {Dukes}, C.~A.,
  {Cipriani}, F., {Witasse}, O., {Schaible}, M.~J., {Zanda}, B., {Strazzulla},
  G., {Baragiola}, R.~A., Jul. 2013. {Paucity of Tagish Lake-like parent bodies
  in the Asteroid Belt and among Jupiter Trojans}. \icarus 225, 517--525.

\bibitem[{{Vilas}(1995)}]{vilas1995turn}
{Vilas}, F., May 1995. {Is the U-B color sufficient for identifying water of
  hydration on solar system bodies?} \icarus 115, 217--218.

\bibitem[{Vilas et~al.(1994)Vilas, Jarvis, and Gaffey}]{vilas1994iron}
Vilas, F., Jarvis, K.~S., Gaffey, M.~J., 1994. Iron alteration minerals in the
  visible and near-infrared spectra of low-albedo asteroids. Icarus 109~(2),
  274--283.

\bibitem[{{Vilas} et~al.(2006){Vilas}, {Smith}, {McFadden}, {Gaffey}, {Larson},
  {Hatch}, and {Jarvis}}]{vilas2006pds}
{Vilas}, F., {Smith}, B.~A., {McFadden}, L.~A., {Gaffey}, M.~J., {Larson},
  S.~M., {Hatch}, E.~C., {Jarvis}, K.~S., Mar. 2006. {Vilas Asteroid Spectra
  V1.1}. NASA Planetary Data System 45.

\bibitem[{{Wong} and {Brown}(2017)}]{2017AJ....153...69W}
{Wong}, I., {Brown}, M.~E., Feb. 2017. {The Color and Magnitude Distribution of
  Hilda Asteroids: Comparison with Jupiter Trojans}. \aj 153, 69.

\end{thebibliography}

	
	\onecolumn
	\begin{deluxetable}{lccccccccr}
		\setcounter{table}{2}
		\tabletypesize{\scriptsize}
		
		\tablecaption{Results for the analysis of the visible parameters in Cybele and Hilda populations for objects observed in SOAR with GHTS.}
		\tablewidth{0pt}
		\tablehead{
			\colhead{Number} & \colhead{0.7 $\mu$m}  & \colhead{0.7 $\mu$m} & \colhead{Turn point} & \colhead{Visible} & \colhead{Visible} & \colhead{Uv} & \colhead{Uv} & \colhead{Taxonomy} & \colhead{Group}\\
			
			\colhead{} & \colhead{depth($\%$)} &  \colhead{Central} & \colhead{($\mu$m)} & \colhead{slope}& \colhead{slope} & \colhead{Slope}& \colhead{Slope}& \colhead{}\\
			
			\colhead{} & \colhead{} &  \colhead{wavelenght($\mu$m)} & \colhead{} &  \colhead{S'\%/1000 Å}& \colhead{unc} & \colhead{S'\%/1000 Å}& \colhead{unc}& \colhead{}
		}
		\startdata
		225  & - & - & 4979.05 $\pm$ 10.96 & -5.62 & 1.25 & -2.907 & 0.592 & B & Cybele \\
		229  & - & - & 5242.28 $\pm$ 7.44 & 3.06 & 0.92 & 14.139 & 0.244 & Xc & Cybele \\
		401  & - & - & -                  & 4.01 & 1.26 & - & - & Cb & Cybele \\
		528  & - & - & - & 1.89 & 0.92 & - & - & Cb & Cybele \\
		790  & - & - & - & 2.77 & 0.42 & - & - & X & Cybele \\
		909  & - & - & - & 2.99 & 1.23 & - & - & X & Cybele \\
		940  & 1.81 $\pm$ 0.01 & 7004.11 $\pm$ 3.82 & 5160.79 $\pm$ 6.24 & 1.96 & 0.53 & 13.337 & 0.414 & Cgh & Cybele \\
		1177 & - & - & - & 2.89 & 0.92 & - & - & X & Cybele \\
		1280 & - & - & 5339.43 $\pm$ 10.70 & 1.55 & 0.42 & 12.909 & 0.247 & C & Cybele \\
		6039 & - & - & - & 9.69 & 0.54 & - & - & D & Cybele \\
		\hline
		334  & - & - & 5251.87 $\pm$ 11.34 & 2.87 & 1.23 & 23.125 & 0.622 & Xc & Hilda \\
		1144 & - & - & - & 11.85 & 0.60 & - & - & D & Hilda \\
		1269 & - & - & - & 9.08 & 0.53 & - & - & D & Hilda \\
		1439 & - & - & - & 3.33 & 0.93 & - & - & X & Hilda \\
		1902 & - & - & - & 2.76 & 0.92 & - & - & X & Hilda \\
		3202 & - & - & - & 13.63 & 0.56 & - & - & D & Hilda \\
		3577 & - & - & - & 9.33 & 0.53 & - & - & D & Hilda \\
		3843 & - & - & - & 5.04 & 1.25 & - & - & X & Hilda \\
		7394 & - & - & - & 7.67 & 0.45 & - & - & D & Hilda \\
		\enddata
		\label{table:vissoar}
	\end{deluxetable}

	\begin{deluxetable}{lcccccccr}
		\tabletypesize{\scriptsize}
		\tablecaption{Results for the analysis of the visible parameters in Cybele population for objects extracted from the literature. The '*' symbol is placed when the wavelenght coverage is not suitable for measuring the determined feature, while the symbol '-' is used for indicating that the wavelengh coverage is appropriated, but the feature was not identified. References: (1) \cite{lagerkvist2005s}; (2) SMASS \cite{smassii};(3) S3OS2 \cite{lazzaro2004s30s2}; (4) \cite{vilas2006pds}}
		\tablewidth{0pt}
		\tablehead{
			\colhead{Number} & \colhead{0.7 $\mu$m}  & \colhead{0.7 $\mu$m} & \colhead{Turn point} & \colhead{Visible}  & \colhead{Uv} & \colhead{Taxonomy} & \colhead{Reference}\\
			
			\colhead{} & \colhead{depth} &  \colhead{Central} & \colhead{($\mu$m)} & \colhead{slope}& \colhead{Slope}& \colhead{}\\
			
			\colhead{} & \colhead{($\%$)} &  \colhead{wavelenght($\mu$m)} & &  \colhead{(S'\%/1000\AA)} & \colhead{(S'\%/1000\AA)}& \colhead{}
		}
		\startdata
	 	65 & - & - & * & 0.97 $\pm$ 1.02 & * & C & 3 \\
& - & - & * & 3.11 $\pm$ 1.07 & * & X & 4 \\
& - & - & 0.664 $\pm$ 0.003 & 0.81 $\pm$ 1.06 & 7.96 $\pm$ 1.56 & Xk & 2 \\
76 & - & - & * & 1.54 $\pm$ 1.02 & * & Cb & 4 \\
& - & - & 0.593 $\pm$ 0.006 & 1.24 $\pm$ 1.02 & 4.62 $\pm$ 1.92 & C & 2 \\
87 & - & - & - & 3.02 $\pm$ 1.04 & - &  Xc & 2 \\
& - & - & * & 3.66 $\pm$ 1.02 & * &  X & 3 \\
107 & - & - & * & 2.01 $\pm$ 1.02 &  * & Cb & 3 \\
& - & - & 0.654 $\pm$ 0.004 & 1.24 $\pm$ 1.06 & 2.45 $\pm$ 2.22 & Xk & 2 \\
121 & 2.20 $\pm$ 0.09 & 0.685 $\pm$ 0.001 & * & 1.66 $\pm$ 1.03 & * & Cgh & 4 \\
& 2.32 $\pm$ 0.05 & 0.719 $\pm$ 0.002 & 0.551 $\pm$ 0.001 & 0.24 $\pm$ 1.06 & 10.99 $\pm$ 2.52 & Cgh & 2 \\
1.68 & 1.83 $\pm$ 0.21 & 0.706 $\pm$ 0.011 & 0.552 $\pm$ 0.005 & -1.19 $\pm$ 1.05 & 3.60 $\pm$ 1.65 & Ch & 2 \\
& 2.36 $\pm$ 0.04 & 0.691 $\pm$ 0.001 & * & 1.01 $\pm$ 1.02 & *  & Cgh & 3 \\
225 & - & - & * & 1.63 $\pm$ 1.05 & * &  Cb & 4 \\
229 & - & - & * & 0.75 $\pm$ 1.02 & * & Cb & 3 \\
260 & - & - & * & 3.11 $\pm$ 1.02 & * & Xc & 3 \\
414 & - & - & 0.6$\pm$0.002 & 0.72 $\pm$ 1.05 & 13.15 $\pm$ 2.45 & Cg & 2 \\
& 1.79 $\pm$ 0.03 & 0.723 $\pm$ 0.001 & * & 1.30 $\pm$ 1.03 & * & Cgh & 3 \\
420 & - & - & - & 6.85 $\pm$ 1.04 & - & D & 1 \\
483 & - & - & * & 9.36 $\pm$ 1.09 & * &  L & 4 \\
522 & - & - & - & 4.46 $\pm$ 1.03 & - & X & 1 \\
& - & - & * & 2.99 $\pm$ 1.02 & * &  X & 3 \\
528 & - & - & * & 1.07 $\pm$ 1.03 & * & Cb & 4 \\
536 & - & - & - & 3.77 $\pm$ 1.05 & - & X & 1 \\
& - & - & * & 2.93 $\pm$ 1.01 & *  & X & 3 \\
566 & - & - & * & 2.79 $\pm$ 1.02 & * & X & 4 \\
570 & - & - & - & 5.62 $\pm$ 1.05 & - & T & 2 \\
& - & - & * & 11.65 $\pm$ 1.04 & * & D & 4 \\
643 & - & - & - & 3.99 $\pm$ 1.05 & - & X & 1 \\
& - & - & * & 4.81 $\pm$ 1.06 & * & X & 4 \\
692 & - & - & * & 4.73 $\pm$ 1.09 & * & S & 4 \\
& - & - & * & 6.12 $\pm$ 1.09 & * & S & 3 \\
713 & - & - & - & 0.97 $\pm$ 1.03 & - & C & 2 \\
& 1.34 $\pm$ 0.06 & 0.765$\pm$ 0.001 & * & 1.42 $\pm$ 1.02 & * & C & 3 \\
721 & - & - & - & 8.79 $\pm$ 1.04 & - & D & 1 \\
& - & - & * & 6.29 $\pm$ 1.03 & * & T & 3 \\
733 & - & - & * & 2.15 $\pm$ 1.02 & * & Cb & 4 \\
790 & - & - & * & 3.87 $\pm$ 1.02 & * & X & 3 \\
940 & 2.35 $\pm$ 0.10 & 0.69 $\pm$ 0.003 & * & 2.51 $\pm$ 1.03 & * & Xc & 4 \\
1004 & - & - & - & 5.70 $\pm$ 1.02 & - & T & 1 \\
& - & - & * & 4.44 $\pm$ 1.02 & * & X & 3 \\
1028 & - & - & * & 1.25 $\pm$ 1.02 & * & Cb & 3 \\
1154 & - & - & * & 3.67 $\pm$ 1.03 & * & X & 3 \\
1167 & - & - & * & 9.19 $\pm$ 1.04 & * & D & 4 \\
1177 & - & - & * & 1.09 $\pm$ 1.03 & * & C & 3 \\
1266 & - & - & - & 4.19 $\pm$ 1.04 & - & Xe & 1 \\
& - & - & * & 6.18 $\pm$ 1.02 & * & T & 3 \\
1280 & - & - & * & 4.14 $\pm$ 1.01 & * & X & 3 \\
1328 & - & - & * & 12.52 $\pm$ 1.04 & * & D & 3 \\
1373 & - & - & - & 3.55 $\pm$ 1.07 & - & Xe & 2 \\
1390 & - & - & * & 5.39 $\pm$ 1.08 & * & T & 4 \\
1467 & 4.65 $\pm$ 0.18 & 0.684 $\pm$ 0.003 & * & 2.69 $\pm$ 1.06 & * & C & 4 \\
& 4.97 $\pm$ 0.03 & 0.705 $\pm$ 0.001 & * & -0.88 $\pm$ 1.04 & * & Ch & 3 \\
1556 & - & - & * & 5.44 $\pm$ 1.02 & * & T & 3 \\
1574 & - & - & - & 9.56 $\pm$ 1.03 & - & D & 1 \\
& - & - & * & 9.72 $\pm$ 1.02 & * & D & 3 \\
1579 & - & - & * & -1.19 $\pm$ 1.02 & * & B & 3 \\
1796 & - & - & * & 1.80 $\pm$ 1.03 & * & Cb & 3 \\
& - & - & 0.63 $\pm$ 0.01 & -0.438 $\pm$ 1.07 & 2.114 $\pm$ 2.389 & C & 2 \\
1841 & - & - & * & 3.08 $\pm$ 1.04 & * & X & 3 \\
2266 & - & - & * & 8.64 $\pm$ 1.02 & * & D & 3 \\
2634 & - & - & - & 5.36 $\pm$ 1.03 & - & T & 1 \\
& - & - & * & 2.90 $\pm$ 1.02 & * & X & 3 \\
2891 & - & - & * & 8.83 $\pm$ 1.02 & * & D & 3 \\
3015 & - & - & - & 4.34 $\pm$ 1.06 & - & X & 1 \\
& - & - & * & 6.59 $\pm$ 1.02 & * & D & 3 \\
3095 & - & - & - & 8.20 $\pm$ 1.04 & - & D & 1 \\
3141 & - & - & * & 9.22 $\pm$ 1.04 & * & D & 3 \\
3622 & - & - & - & 10.10 $\pm$ 1.04 & - & D & 1 \\
3675 & - & - & - & 3.67 $\pm$ 1.08 & - & S & 1 \\
4003 & - & - & - & 7.09 $\pm$ 1.06 & - & L & 1 \\
4158 & - & - & - & 4.69 $\pm$ 1.05 & - & T & 1 \\
4973 & - & - & - & 8.31 $\pm$ 1.04 & - & D & 1 \\
5301 & 4.346$\pm$0.065 & 0.759$\pm$0.002 & * & 1.19 $\pm$ 1.07 & * & Ch & 3 \\
5362 & - & - & * & 6.69 $\pm$ 1.04 & * & T & 3 \\
5780 & - & - & 0.559$\pm$0.005 & 1.79 $\pm$ 1.10 & 22.55 $\pm$ 2.16 & C & 1 \\
5833 & - & - & - & 5.16 $\pm$ 1.07 & - & X & 1 \\
5914 & - & - & * & 7.88 $\pm$ 1.03 & * & D & 3 \\
6057 & - & - & * & 2.50 $\pm$ 1.05 & * & Xc & 3 \\
		\enddata
		\label{table:viscybele}
	\end{deluxetable}

	\begin{deluxetable}{lccccccccr}
		\tabletypesize{\scriptsize}
		\tablecaption{Results for the analysis of the visible parameters in Hilda population for objects extracted from the literature. The '*' symbol is placed when the wavelenght coverage are not suitable for measuring the determined feature, while the symbol '-' is used for indicating that the wavelengh coverage is appropriated, but the feature was not identified.. References: (1) \cite{dahlgren1997s} and \cite{dahlgren1997s}; (2) SMASS \citep{smassii};(3) S3OS2 \citep{lazzaro2004s30s2}; (4) \cite{vilas2006pds}}
		\tablewidth{0pt}
		\tablehead{
			\colhead{Number} & \colhead{0.7 $\mu$m}  & \colhead{0.7 $\mu$m} & \colhead{Turn point} & \colhead{Visible}  & \colhead{Uv} & \colhead{Taxonomy} & \colhead{Reference}\\
			
			\colhead{} & \colhead{depth} &  \colhead{Central} & \colhead{($\mu$m)} & \colhead{slope}& \colhead{Slope}& \colhead{}\\
			
			\colhead{} & \colhead{($\%$)} &  \colhead{wavelenght($\mu$m)} & &  \colhead{(S'\%/1000\AA)} & \colhead{(S'\%/1000\AA)}& \colhead{}
		}
		\startdata
153 & - & - & - & 2.08 $\pm$ 1.04 & - & X & 2 \\
& - & - & - & 3.02 $\pm$ 1.04 & - &  X & 1 \\
& - & - & * & 3.77 $\pm$ 1.03 & * & X & 4 \\
190 & - & - & - & 1.58 $\pm$ 1.04 & - & Xc & 2 \\
& - & - & - & 2.83 $\pm$ 1.02 & - & X & 1 \\
334 & - & - & - & 2.37 $\pm$ 1.04 & - & C & 1 \\
& - & - & * & 2.79 $\pm$ 1.03 & * & Xc & 4 \\
361 & - & - & - & 6.23 $\pm$ 1.02 & - & T & 1 \\
& - & - & * & 6.57 $\pm$ 1.02 & * & D & 3 \\
449 & - & - & - & 3.14 $\pm$ 1.05 & - & X & 1 \\
748 & - & - & * & 4.82 $\pm$ 1.05 & * & T & 4 \\
958 & - & - & - & 8.19 $\pm$ 1.02 & - & D & 1 \\
1038 & - & - & - & 8.56 $\pm$ 1.05 & - & D & 1 \\
& - & - & - & 8.96 $\pm$ 1.04 & - & D & 1 \\
1162 & - & - & * & 3.66 $\pm$ 1.04 & * & X & 4 \\
1180 & - & - & * & 4.78 $\pm$ 1.02 & * & X & 3 \\
1202 & - & - & - & 8.42 $\pm$ 1.03 & - & D & 1 \\
1212 & - & - & - & 2.71 $\pm$ 1.06 & - & X & 2 \\
& - & - & - & 5.35 $\pm$ 1.08 & - & X & 1 \\
& - & - & - & 6.30 $\pm$ 1.06 & - & T & 1 \\
1268 & - & - & - & 6.35 $\pm$ 1.05 & - & T & 1 \\
& - & - & - & 10.69 $\pm$ 1.08 & - & D & 1 \\
1345 & - & - & - & 3.03 $\pm$ 1.07 & - & Xc & 1 \\
1439 & - & - & - & 2.12 $\pm$ 1.09 & - & C & 1 \\
1512 & - & - & * & 5.01 $\pm$ 1.03 & * & X & 4 \\
1529 & - & - & - & 9.25 $\pm$ 1.05 & - & D & 1 \\
& - & - & - & 9.67 $\pm$ 1.03 & - & D & 1 \\
1754 & - & - & * & 3.76 $\pm$ 1.02 & * & X & 3 \\
2246 & - & - & - & 6.73 $\pm$ 1.07 & - & D & 2 \\
2483 & - & - & - & 11.88 $\pm$ 1.05 & - & D & 1 \\
2959 & - & - & * & 9.92 $\pm$ 1.07 & * & D & 3 \\
& - & - & - & 10.53 $\pm$ 1.05 & - & D & 1 \\
3134 & - & - & - & 8.13 $\pm$ 1.02 & - & D & 1 \\
3254 & - & - & - & 5.55 $\pm$ 1.16 & - & D & 2 \\
3415 & - & - & - & 9.72 $\pm$ 1.06 & - & D & 1 \\
& - & - & - & 11.63 $\pm$ 1.09 & - & D & 1 \\
3514 & - & - & - & 9.37 $\pm$ 1.07 & - & D & 1 \\
3561 & - & - & - & 8.01 $\pm$ 1.06 & - & D & 1 \\
3655 & - & - & - & 9.10 $\pm$ 1.07 & - & D & 1 \\
3694 & - & - & - & 10.68 $\pm$ 1.04 & - & D & 1 \\
3843 & - & - & - & 3.49 $\pm$ 1.06 & - & X & 1 \\
& - & - & - & 4.43 $\pm$ 1.06 & - & X & 1 \\
3923 & - & - & - & 3.24 $\pm$ 1.07 & - & X & 1 \\
& - & - & - & 6.08 $\pm$ 1.07 & - & T & 1 \\
3990 & - & - & * & 10.80 $\pm$ 1.03 & * & D & 3 \\
		\enddata
		\label{table:vishilda}
	\end{deluxetable}
	
\begin{deluxetable}{lrrrrc}
	\tabletypesize{\scriptsize}
	\tablecaption{Results for Hildas near-IR spectra obseved with TNG}
	\tablewidth{0pt}
	\tablehead{
		\colhead{Number} & \colhead{IR Slope} & \colhead{IR Slope} & \colhead{mIR Slope} & \colhead{mIR Slope}& \colhead{Taxonomy} \\
		\colhead{ } & \colhead{S'\%/1000 Å} & \colhead{unc} & \colhead{S'\%/1000 Å} & \colhead{unc}& \colhead{} }
	\startdata
	190	  & 0.886 &	1.052 &	1.138  & 1.447  & Cg \\
	334	  & 1.835 &	1.041 &	-0.268 & 1.173  & X  \\	
	1202  & 4.044 & 1.065 & 2.068  & 1.502  & D  \\
	1269  &	4.461 &	1.081 &	-0.453 & 1.234  & D  \\
	1754  &	2.193 &	1.029 &	2.825  & 1.387  & X  \\
	2067  &	4.123 &	1.148 &	3.225  & 1.226  & D  \\
	& 4.561 & 1.078 & 0.201  & 1.479  & D  \\
	2624  &	5.259 &	1.103 &	-0.364 & 1.435  & D  \\
	3557  &	4.246 & 1.065 &	0.965  & 1.548  & D  \\
	3561  &	4.176 &	1.086 &	1.871  & 1.436  & D  \\
	4317  &	5.013 &	1.087 &	2.725  & 1.579  & D  \\
	5368  &	4.675 &	1.073 &	0.873  & 1.552  & D  \\
	5661  &	4.574 &	1.111 &	0.725  & 1.356  & D  \\
	5711  &	5.861 &	1.109 &	3.115  & 1.676  & D  \\
	6237  &	2.652 &	1.101 &	-1.93  & 1.773  & X  \\
	9121  &	4.488 &	1.083 &	2.715  & 1.318  & D  \\
	11750 &	4.411 &	1.078 &	3.025  & 1.356  & D  \\
	15505 & 4.471 & 1.071 & 2.469  & 1.681  & D  \\
	15417 &	5.160 &	1.202 &	-0.24  & 1.712  & D  \\
	15540 &	4.022 &	1.110 &	6.012  & 1.797  & D  \\
	\enddata
	\label{table:hildasir}
\end{deluxetable}

	\begin{deluxetable}{lrrrrccc}
		\tabletypesize{\scriptsize}
		\tablecaption{Results for near-IR parametrization for objects extracted from the literature. Reference:(1) \cite{reddypds}; (2) SMASS II \citep{smassii}; (3) \cite{takir2012outer}}
		\tablewidth{0pt}
		\tablehead{
			\colhead{Number} & \colhead{IR Slope} & \colhead{IR Slope} & \colhead{mIR Slope} & \colhead{mIR Slope}& \colhead{Taxonomy} & \colhead{Group} &\colhead{Reference} \\
			\colhead{ } & \colhead{S'\%/1000 Å} & \colhead{unc} & \colhead{S'\%/1000 Å} & \colhead{unc}& \colhead{}& \colhead{} }
		\startdata
		76  & 2.197 & 1.013 & 1.835 & 1.013 & X & Cybele& 3\\
		    & 1.838 & 1.152 & 2.344 & 1.243 & X & & 2\\
		87  & 1.038 & 1.013 & 1.771 & 1.021 & Xc & Cybele& 1\\
		107 & 1.229 & 1.012 & 1.647 & 1.012 & C & Cybele& 3\\
		121 & 0.778 & 1.016 & 0.164 & 1.022 & L & Cybele& 1\\
		    & 2.001 & 1.011 & -0.134 & 1.022 & K & & 2\\
		    & 1.850 & 1.014 & 0.030 & 1.023 & K & & 3\\
		401 & 2.226 & 1.022 & 1.614 & 1.033 & X & Cybele& 3\\
		790 & 1.931 & 1.022 & 1.451 & 1.040 & X & Cybele& 3\\
		\hline
		153 & 2.221 & 1.010 & 1.782 & 1.042 & X &Hilda& 2 \\
		    & 2.240 & 1.012 & 1.813 & 1.034 & X & &3 \\
		190 & 1.823 & 1.016 & 1.605 & 1.046 & X &Hilda& 3 \\
		334 & 1.397 & 1.009 & 1.551 & 1.068 & Xc &Hilda& 2 \\
		    & 1.580 & 1.018 & 1.545 & 1.118 & X & & 3 \\
		361 & 3.156 & 1.021 & 2.623 & 1.074 & X & Hilda& 3 \\
		
		\enddata
		\label{table:liteir}
	\end{deluxetable}

	\begin{deluxetable}{l l c c c l c c c c c c c c}
		\tabletypesize{\scriptsize}
		\tablecaption{Cybele properties table}
		\tablewidth{0pt}
		\tablehead{\colhead{Number} & \colhead{Name} &\colhead{$a_p$} & \colhead{$e_p$}  & \colhead{$\sin{i_p}$} &\colhead{Family}& \colhead{$H$} &  \colhead{$p_V$} & \colhead{$p_{V err}$} &  \colhead{$D$} & \colhead{$D_{err}$} \\ 
		\colhead{} & \colhead{} & \colhead{($au$)} & \colhead{}  & \colhead{} & \colhead{} & \colhead{(mag)} & \colhead{} & \colhead{} & \colhead{($km$)} & \colhead{($km$)}} 
		
		\startdata
			65 &  Cibele  & 3.429 & 0.111 & 3.563 &  -   &  6.62  & 0.059 & 0.039 & 276.584 & 74.487 \\
76 &  Freia  & 3.411 & 0.166 & 2.122 &  -  &  7.90  & 0.058 & 0.004 & 145.423 & 1.287 \\
87 &  Sylvia  & 3.485 & 0.054 & 0.171 &  Sylvia  &  6.94 & 0.046 & 0.004 & 253.051 & 2.953 \\
107 &  Camilla  & 3.486 & 0.093 & 0.169 &  Sylvia  &  7.08  & 0.059 & 0.012 & 210.37 & 8.326 \\
121 &  Hermione  & 3.447 & 0.134 & 7.598 &  -  & 7.31  & 0.076 & 0.034 & 166.242 & 8.807 \\
168 &  Sibylla  & 3.379 & 0.072 & 4.666 &  -  &  7.94  & 0.056 & 0.012 & 145.366 & 3.219 \\
225 &  Henrietta  & 3.389 & 0.264 & 20.873 &  -  &  8.72  & 0.062 & 0.008 & 95.934 & 1.249 \\
229 &  Adelinda  & 3.421 & 0.139 & 2.079 &  -  & 9.13  & 0.035 & 0.007 & 105.912 & 1.779 \\
260 &  Huberta  & 3.444 & 0.115 & 6.416 &  -  &  8.97  & 0.044 & 0.01 & 101.539 & 0.941 \\
401 &  Ottilia  & 3.346 & 0.036 & 5.972 &  -  &  9.20  & 0.052 & 0.009 & 87.803 & 0.435 \\
414 &  Liriope  & 3.504 & 0.072 & 9.558 &  -  &  9.49  & 0.027 & 0.003 & 88.76 & 2.169 \\
420 &  Bertholda  & 3.417 & 0.031 & 6.687 &  -  &  8.40  & 0.044 & 0.004 & 138.699 & 3.446 \\
522 &  Helga  & 3.63 & 0.085 & 4.417 &  -  & 9.00  & 0.057 & 0.0133 & 83.7 & 4.85 \\
528 &  Rezia  & 3.403 & 0.018 & 12.685 &  -  &  9.14  & 0.046 & 0.006 & 91.966 & 0.361 \\
536 &  Merapi  & 3.499 & 0.086 & 19.424 &  -  &  8.2  & 0.048 & 0.005 & 147.066 & 5.524 \\
570 &  Kythera  & 3.426 & 0.12 & 1.788 &  -  &  8.81 & 0.069 & 0.004 & 87.486 & 0.784 \\
643 &  Scheherezade  & 3.361 & 0.058 & 13.769 &  -  &  9.70  & 0.058 & 0.013 & 64.997 & 0.382 \\
692 &  Hippodamia  & 3.383 & 0.17 & 26.079 &  -  & 9.18 & 0.205 & 0.029 & 42.771 & 0.633 \\
713 &  Luscinia  & 3.392 & 0.164 & 10.36 &  -  &  8.97  & 0.048 & 0.005 & 97.968 & 0.876 \\
721 &  Tabora  & 3.55 & 0.116 & 8.323 &  -  &  9.26  & 0.048 & 0.006 & 74.791 & 0.525 \\
790 &  Pretoria  & 3.412 & 0.151 & 20.527 &  -  & 8.00  & 0.041 & 0.029 & 163.4 & 53.372 \\
909 &  Ulla  & 3.543 & 0.05 & 0.308 &  Ulla  &  8.95 & 0.037 & 0.001 & 113.13 & 1.48 \\
940 &  Kordula  & 3.376 & 0.172 & 6.21 &  -  &  9.55  & 0.041 & 0.009 & 79.852 & 0.504 \\
1004 &  Belopolskya  & 3.402 & 0.087 & 2.979 &  -  &  9.99  & 0.028 & 0.001 & 79.83 & 1.33 \\
1028 &  Lydina  & 3.408 & 0.107 & 9.393 &  -  &  9.43  & 0.038 & 0.006 & 88.526 & 0.762 \\
1154 &  Astronomia  & 3.39 & 0.071 & 4.533 &  -  & 10.51 & 0.036 & 0.008 & 55.715 & 0.5 \\
1177 &  Gonnessia  & 3.349 & 0.031 & 15.069 &  -  & 9.66 & 0.032 & 0.016 & 104.631 & 33.728 \\
1266 &  Tone  & 3.359 & 0.051 & 17.185 &  -  &  9.41  & 0.053 & 0.005 & 75.47 & 0.523 \\
1280 &  Baillauda  & 3.415 & 0.05 & 6.459 &  -  & 9.99  & 0.045 & 0.001 & 53.97 & 0.72 \\
1328 &  Devota  & 3.506 & 0.135 & 5.765 &  -  & 10.09 & 0.046 & 0.005 & 53.697 & 0.481 \\
1373 &  Cincinnati  & 3.422 & 0.314 & 38.u929 &  -  & 11.37 & 0.155 & 0.036 & 19.448 & 0.175 \\
1467 &  Mashona  & 3.384 & 0.131 & 21.947 &  -  &  8.57 & 0.083 & 0.014 & 89.16 & 0.728 \\
1556 &  Wingolfia  & 3.427 & 0.109 & 15.748 &  -  & 10.67 &  0.093 & 0.012 & 33.88 & 2.12 \\
1574 &  Meyer  & 3.537 & 0.035 & 14.478 &  -  &  9.90  & 0.042 & 0.011 & 57.785 & 0.435 \\
1579 &  Herrick  & 3.437 & 0.127 & 8.762 &  -  & 10.77 & 0.043 & 0.006 & 46.925 & 0.405 \\
1796 &  Riga  & 3.356 & 0.057 & 22.585 &  -  &  9.84  & 0.044 & 0.005 & 68.167 & 0.298 \\
1841 &  Masaryk  & 3.422 & 0.1 & 2.62 &  -  & 10.94 &  0.052 & 0.005 & 40.24 & 0.504 \\
2266 &  Tchaikovsky  & 3.4 & 0.182 & 13.247 &  -  & 10.88 & 0.045 & 0.002 & 43.58 & 0.69 \\
2634 &  James Bradley  & 3.457 & 0.049 & 6.422 &  -  &  10.50  &  0.107 & 0.005 & 33.726 & 0.488 \\
2891 &  McGetchin  & 3.355 & 0.136 & 9.296 &  -  &  11.0  &  0.061 & 0.005 & 33.996 & 0.418 \\
3015 &  Candy  & 3.385 & 0.173 & 17.403 &  -  & 11.145 & 0.107 & 0.017 & 24.517 & 0.47 \\
3095 &  Omarkhayyam  & 3.502 & 0.075 & 2.966 &  -  & 10.949 & 0.063 & 0.009 & 29.007 & 0.335 \\
3141 &  Buchar  & 3.4 & 0.077 & 10.995 &  -  &  10.80  & 0.043 & 0.004 & 29.368 & 0.231 \\
3622 &  Ilinsky  & 3.389 & 0.043 & 4.935 &  -  &  11.80  & 0.102 & 0.023 & 21.88 & 0.458 \\
3675 &  Kemstach  & 3.369 & 0.088 & 10.857 &  -  &  11.10  & 0.181 & 0.018 & 18.825 & 0.184 \\
4003 &  Schumann  & 3.427 & 0.094 & 5.059 &  -  &  11.30  & 0.054 & 0.009 & 35.139 & 0.286 \\
4158 &  Santini  & 3.401 & 0.019 & 6.17 &  -  & 11.60  & 0.172 & 0.013 & 16.797 & 0.181 \\
4973 &  Showa  & 3.426 & 0.077 & 18.924 &  -  & 11.50 & 0.068 & 0.01 & 27.958 & 0.423 \\
5301 &  Novobranets  & 3.362 & 0.102 & 10u.047 &  -  &  12.10  & 0.058 & 0.011 & 20.97 & 0.298 \\
5362 &  1978 CH  & 3.389 & 0.024 & 6.146 &  -  &  11.70  & 0.085 & 0.013 & 21.865 & 0.253 \\
5780 &  Lafontaine  & 3.346 & 0.131 & 8.677 &  -  & 12.365 & 0.055 & 0.004 & 22.593 & 0.119 \\
5833 &  Peterson  & 3.491 & 0.032 & 19.381 &  -  & 11.587 & 0.105 & 0.021 & 27.077 & 0.435 \\
5914 &  Kathywhaler  & 3.543 & 0.069 & 0.162 &  Sylvia  & 11.283 & 0.062 & 0.01 & 38.097 & 0.224 \\
6039 &  Parmenides  & 3.411 & 0.057 & 13.11 &  -  &11.90  & 0.076 & 0.004 & 22.03 & 0.157 \\
6057 &  Robbia  & 3.329 & 0.1 & 17.863 &  -  &  11.90  & 0.043 & 0.004 & 29.368 & 0.231 \\
		
		\enddata
	\label{tab:cybprop}
	\end{deluxetable}
	
\begin{deluxetable}{l l c c c l c c c c c c}
		\tabletypesize{\scriptsize}
		\tablecaption{Hildas properties table}
		\tablewidth{0pt}
		\tablehead{\colhead{Number} & \colhead{Name} &\colhead{$a_p$} & \colhead{$e_p$}  & \colhead{$\sin{i_p}$} &\colhead{Family}& \colhead{$H$} &  \colhead{$p_V$} & \colhead{$p_{V err}$} &  \colhead{$D$} & \colhead{$D_{err}$} \\ 
	\colhead{} & \colhead{} & \colhead{($au$)} & \colhead{}  & \colhead{} & \colhead{} & \colhead{(mag)} & \colhead{} & \colhead{} & \colhead{($km$)} & \colhead{($km$)}} 

		\startdata
			153 & Hilda & 3.965 & 0.174 & 0.155 & Hilda & 7.67 & 0.038 & 0.016 & 218.844 & 3.637 \\ 
190 & Ismene & 3.986 & 0.166 & 6.177 & - & 7.59 & 0.035 & 0.001 & 214.664 & 8.608 \\ 
334 & Chicago & 3.895 & 0.022 & 4.641 & -  & 7.70 & 0.041 & 0.013 & 198.77 & 5.668 \\ 
361 & Bononia & 3.96 & 0.214 & 12.626 & -  & 8.22 & 0.038 & 0.008 & 154.334 & 2.69 \\ 
449 & Hamburga & 2.551 & 0.173 & 3.085 & - & 9.47 & 0.033 & 0.009 & 80.827 & 17.911 \\ 
748 & Simeisa & 3.944 & 0.188 & 2.259 & - & 9.01 & 0.041 & 0.007 & 103.725 & 1.034 \\ 
958 & Asplinda & 3.986 & 0.186 & 5.63 & - & 10.49 & 0.045 & 0.005 & 45.117 & 0.091 \\ 
1038 & Tuckia & 3.965 & 0.164 & 0.143 & Hilda & 10.60 & - & - & - & - \\ 
1144 & Oda & 3.748 & 0.094 & 9.743 & - & 10.00 & 0.061 & 0.014 & 56.347 & 0.194 \\ 
1162 & Larissa & 3.93 & 0.109 & 1.887 & - & 9.42 & 0.169 & 0.012 & 42.243 & 0.111 \\ 
1180 & Rita & 3.985 & 0.158 & 7.199 & - & 9.14 & 0.058 & 0.009 & 82.308 & 0.418 \\ 
1202 & Marina & 3.996 & 0.166 & 3.334 & - & 10.09 & - & - & - & - \\ 
1212 & Francette & 3.967 & 0.23 & 0.126 & Hilda & 9.54 & 0.046 & 0.007 & 76.395 & 0.155 \\ 
1268 & Libya & 3.975 & 0.102 & 4.427 & - & 9.12 & 0.043 & 0.003 & 96.708 & 0.848 \\ 
1269 & Rollandia & 3.906 & 0.1 & 2.758 & - & 8.82 & 0.048 & 0 & 104.893 & 0.624 \\ 
1345 & Potomac & 3.989 & 0.183 & 11.399 & - & 9.73 & 0.043 & 0.008 & 72.975 & 0.463 \\ 
1439 & Vogtia & 4.003 & 0.118 & 4.203 & -  & 10.45 & 0.046 & 0.007 & 50.542 & 0.148 \\ 
1512 & Oulu & 3.967 & 0.147 & 6.491 & - & 9.62 & 0.038 & 0.005 & 79.222 & 0.241 \\ 
1529 & Oterma & 3.964 & 0.154 & 0.137 & Hilda & 10.05 & 0.054 & 0.003 & 56.327 & 0.285 \\ 
1754 & Cunningham & 3.941 & 0.169 & 12.153 & - & 9.77 & - & - & - & - \\ 
1902 & Shaposhnikov & 3.965 & 0.222 & 12.496 & - & 9.51 & 0.04 & 0.012 & 83.443 & 1.723 \\ 
2067 & Aksnes & 3.964 & 0.182 & 3.08 & - & 10.55 & 0.054 & 0.003 & 46.003 & 0.761 \\ 
2246 & Bowell & 3.958 & 0.094 & 6.495 & - & 10.56 & 0.045 & 0.012 & 48.424 & 0.429 \\ 
2483 & Guinevere & 3.972 & 0.278 & 4.499 & - & 10.90 & 0.067 & 0.011 & 35.687 & 0.18 \\ 
2624 & Samitchell & 3.948 & 0.117 & 2.797 & - & 10.80 & - & - & - & - \\ 
2959 & Scholl & 3.943 & 0.275 & 5.234 & - & 11.10 & 0.054 & 0.015 & 32.783 & 0.319 \\ 
3134 & Kostinsky & 3.966 & 0.184 & 0.156 & Hilda & 10.50 & 0.037 & 0.004 & 50.389 & 0.403 \\ 
3202 & Graff & 3.936 & 0.115 & 11.107 & - & 11.311 & 0.055 & 0.013 & 35.914 & 0.244 \\ 
3254 & Bus & 3.951 & 0.165 & 4.446 & - & 11.20 & 0.073 & 0.002 & 31.104 & 0.895 \\ 
3415 & Danby & 3.963 & 0.249 & 1.367 & - & 11.304 & 0.063 & 0.006 & 36.582 & 0.124 \\ 
3514 & Hooke & 3.954 & 0.191 & 3.505 & - & 11.70 & 0.084 & 0.012 & 22.037 & 0.073 \\ 
3557 & Sokolsky & 4.003 & 0.173 & 6.049 & - & 10.90 & - & - & - & - \\ 
3561 & Devine & 3.962 & 0.133 & 0.149 & Hilda  & 11.10 & - & - & - & - \\ 
3577 & Putilin & 3.948 & 0.197 & 3.741 & - & 10.56 & 0.051 & 0.003 & 49.138 & 0.313 \\ 
3655 & Eupraksia & 4.014 & 0.2 & 3.823 & - & 11.13 & 0.063 & 0.01 & 36.66 & 0.207 \\ 
3694 & Sharon & 3.933 & 0.206 & 4.976 & - & 10.50 & 0.058 & 0.004 & 46.036 & 0.345 \\ 
3843 & OISCA & 3.993 & 0.144 & 3.926 & - & 10.94 & 0.108 & 0.023 & 30.768 & 0.3 \\ 
3923 & Radzievskij & 3.966 & 0.196 & 0.05 & Schubart  & 11.60 & 0.05 & 0.005 & 29.87 & 0.163 \\ 
3990 & Heimdal & 3.965 & 0.168 & 0.167 & Hilda & 10.90 & 0.067 & 0.021 & 35.679 & 0.33 \\ 
4317 & Garibaldi & 3.967 & 0.213 & 0.159 & Hilda & 10.90 & 0.052 & 0.01 & 38.611 & 0.224 \\ 
5368 & Vitagliano & 3.974 & 0.083 & 6.262 & - & 11.2 & 0.058 & 0.017 & 34.812 & 0.061 \\ 
5661 & Hildebrand & 3.966 & 0.234 & 13.311 & - & 11.10 & - & - & - & - \\ 
5711 & Eneev & 3.942 & 0.164 & 6.371 & - & 11.10 & - & - & - & - \\ 
6237 & Chikushi & 3.935 & 0.073 & 5.362 & - & 11.50 & - & - & - & - \\ 
7394 & Xanthomalitia & 3.933 & 0.033 & 8.61 & - & 11.57 & 0.061 & 0.006 & 32.472 & 0.125 \\ 
9121 & Stefanovalentini & 3.885 & 0.041 & 4.647 & - & 11.30 & - & - & - & - \\ 
11750 & 1999 NM33 & 3.981 & 0.053 & 2.678 & - & 12.40 & 0.07 & 0.007 & 18.244 & 0.336 \\ 
15505 & 1999 RF56 & 3.966 & 0.179 & 0.144 & Hilda & 11.76 & 0.079 & 0.008 & 24.789 & 0.38 \\ 
15417 & Babylon & 3.933 & 0.053 & 3.185 & - & 11.80 & - & - & - & - \\ 
15540 & 2000 CF18 & 3.989 & 0.113 & 16.988 & - & 12.20 &  0.08 & 0.008 & 19.528 & 0.39 \\
		\enddata
	\label{tab:hildaprop}
	\end{deluxetable}
	

	\clearpage


\end{document}